
\input epsf

%
                              %
%
%
\def\and{{\it\&}}
\def\half{{1\over2}}

\def\quarter{{1\over4}}

\def\gesim{\,{\raise-3pt\hbox{$\sim$}}\!\!\!\!\!{\raise2pt\hbox{$>$}}\,}
\def\lesim{\,{\raise-3pt\hbox{$\sim$}}\!\!\!\!\!{\raise2pt\hbox{$<$}}\,}
\def\boldoverdot{\,{\raise6pt\hbox{\bf.}\!\!\!\!\>}}
\def\re{{\bf Re}}
\def\im{{\bf Im}}
\def\ie{{\it i.e.}}

\def\ibid{{\it ibid.}}
\def\etal{{\it et. al.}}
\def\acal{{\cal A}}

\def\lcal{{\cal L}}
\def\mcal{{\cal M}}
\def\ncal{{\cal N}}
\def\ocal{{\cal O}}
\def\pcal{{\cal P}}

\def\ZZ{{\bf Z}}

\def\xibf{{\pmb{$\xi$}}}

\def\vev{vacuum expectation value}

\def\diag{\hbox{\diag}}
\def\sm{Standard Model}

\def\ev{~\hbox{eV}}

\def\gev{~\hbox{GeV}}
\def\tev{~\hbox{TeV}}

%
                               %
%
%
\def\doubleundertext#1{
{\undertext{\vphantom{g}#1}}\par\nobreak\vskip-\the\baselineskip\vskip 2pt%
\noindent\ \undertext{\phantom{#1} \hbox to .5in{}}}
\def\noblackbox{\overfullrule=0pt}
\def\leaderfill{\leaders\hbox to 1 em{\hss.\hss}\hfill}
\def\expel{\par\vfill\eject}
\def\inbox#1{\vbox{\hrule\hbox{\vrule\kern5pt
     \vbox{\kern5pt#1\kern5pt}\kern5pt\vrule}\hrule}}
\def\sqr#1#2{{\vcenter{\hrule height.#2pt
      \hbox{\vrule width.#2pt height#1pt \kern#1pt
         \vrule width.#2pt}
      \hrule height.#2pt}}}
\def\square{\mathchoice\sqr56\sqr56\sqr{2.1}3\sqr{1.5}3}
\def\today{\ifcase\month\or
  January\or February\or March\or April\or May\or June\or
  July\or August\or September\or October\or November\or December\fi
  \space\number\day, \number\year}
\def\pmb#1{\setbox0=\hbox{#1}%
  \kern-.025em\copy0\kern-\wd0
  \kern.05em\copy0\kern-\wd0
  \kern-.025em\raise.0433em\box0 }
\def\up#1{^{\left( #1 \right) }}
\def\lowti#1{_{{\rm #1 }}}
\def\inv#1{{1\over#1}}
\def\su#1{{SU(#1)}}
\def\ui{U(1)}
\def\antes{}
\def\despues{.}
%

%
\def\sumprime_#1{\setbox0=\hbox{$\scriptstyle{#1}$}
  \setbox2=\hbox{$\displaystyle{\sum}$}
  \setbox4=\hbox{${}'\mathsurround=0pt$}
  \dimen0=.5\wd0 \advance\dimen0 by-.5\wd2
  \ifdim\dimen0>0pt
  \ifdim\dimen0>\wd4 \kern\wd4 \else\kern\dimen0\fi\fi
\mathop{{\sum}'}_{\kern-\wd4 #1}}
\def\sumbiprime_#1{\setbox0=\hbox{$\scriptstyle{#1}$}
  \setbox2=\hbox{$\displaystyle{\sum}$}
  \setbox4=\hbox{${}'\mathsurround=0pt$}
  \dimen0=.5\wd0 \advance\dimen0 by-.5\wd2
  \ifdim\dimen0>0pt
  \ifdim\dimen0>\wd4 \kern\wd4 \else\kern\dimen0\fi\fi
\mathop{{\sum}''}_{\kern-\wd4 #1}}
\def\sumtriprime_#1{\setbox0=\hbox{$\scriptstyle{#1}$}
  \setbox2=\hbox{$\displaystyle{\sum}$}
  \setbox4=\hbox{${}'\mathsurround=0pt$}
  \dimen0=.5\wd0 \advance\dimen0 by-.5\wd2
  \ifdim\dimen0>0pt
  \ifdim\dimen0>\wd4 \kern\wd4 \else\kern\dimen0\fi\fi
\mathop{{\sum}'''}_{\kern-\wd4 #1}}
%
%
\newcount\chapnum
\def\clearchap{\chapnum=0}
\def\chap#1{\clearsect\clearprob
\global\advance\chapnum by 1 \par\vskip .5 in\par%
\centerline{{\bigboldiii\antes\the\chapnum\despues\ #1}}}
\newcount\sectnum
\def\clearsect{\sectnum=0}
\def\sect#1{\clearprob\global\advance\sectnum by 1 \par\vskip .25 in\par%
\noindent{\bigboldii\the\chapnum.\the\sectnum:\ #1}\nobreak}
\newcount\yesnonum
\def\clearyesno{\yesnonum=0}
\def\verify{\global\advance\yesnonum by 1{\bf (VERIFY!!)}}
\def\tocheck{\par\vskip 1 in{\bigboldv TO VERIFY: \the\yesnonum\ ITEMS.}}
\newcount\notenum

\def\noteeye{%
\hbox{{$\quad(\!(\subset\!\!\!\!\bullet\!\!\!\!\supset)\!)\quad$}}}

\def\note#1{\global\advance\notenum by 1{\bf \noteeye #1 \noteeye } }
\def\noteout{\par\vskip 1 in{\bigboldiv NOTES: \the\notenum.}}
\newcount\borrownum

\def\borrow{\global\advance\borrownum by 1{\bigboldi BORROWED BY:\ }}
\def\borrowed{\par\vskip 0.5 in{\bigboldii BOOKS OUT:\ \the\borrownum.}}
\newcount\refnum

\def\ref#1{\global\advance\refnum by 1\item{\the\refnum.\ }#1}
\def\stariref#1{\global\advance\refnum by 1\item{%
               {\bigboldiv *}\the\refnum.\ }#1}
\def\stariiref#1{\global\advance\refnum by 1\item{%
               {\bigboldiv **}\the\refnum.\ }#1}
\def\stariiiref#1{\global\advance\refnum by 1\item{
               {\bigboldiv ***}\the\refnum.\ }#1}
\newcount\probnum
\def\clearprob{\probnum=0}
\def\prob{\global\advance\probnum by 1 {\medskip $\triangleright$\
\undertext{{\sl Problem}}\ \the\chapnum.\the\sectnum.\the\probnum.\ }}
\newcount\probchapnum

\def\probchap{\global\advance\probchapnum by 1 {\medskip $\triangleright$\
\undertext{{\sl Problem}}\ \the\chapnum.\the\probchapnum.\ }}
\def\undertext#1{$\underline{\smash{\hbox{#1}}}$}
%

%
%
%

\def\UCR{
\address
{{\it Department of Physics\break
                  University of California at Riverside\break
                  Riverside, California 92521--0413; U.S.A. \break
                  \bit}}}

\def\bit{{E{\rm-}Mail address{\rm:} jose.wudka{\rm@}ucr{\rm.}edu}}
\catcode`\@=11 
%
%
                                %
%
%


%

%

%
\font\bigboldi=cmbx10 scaled\magstep1
\font\bigboldii=cmbx10 scaled\magstep2
\font\bigboldiii=cmbx10 scaled\magstep3
\font\bigboldiv=cmbx10 scaled\magstep4
\font\bigboldv=cmbx10 scaled\magstep5
\font\small=cmr8
\font\smalli=cmr8 scaled\magstep1
\font\smallii=cmr8 scaled\magstep2

\font\smallv=cmr8 scaled\magstep5
\font\eightrm=cmr8
%
%
                     %
%
%
\clearchap
\clearyesno
\headline={\ifnum\pageno>0   {\smalli \title (\today)} \hfil {\small Page
\folio } \else\hfil\fi}
%
%
%
%
\newdimen\fullhsize
\newdimen\fullvsize
\newbox\leftcolumn
\def\fullline{\hbox to\fullhsize}
\gdef\twocol{\fullhsize=9.75in
\hsize=4.5in
\vsize=7in
\advance\hoffset by -.5 in
\def\makeheadline{\vbox to 0pt{\vskip-.4in
  \fullline{\vbox to8.5pt{}\the\headline}\vss}
   \nointerlineskip}
\def\makefootline{\baselineskip=24pt
    \fullline{\the\footline}}
\let\lr=L
\output{\if L\lr
   \global\setbox\leftcolumn=\columnbox \global\let\lr=R
  \else \doubleformat \global\let\lr=L\fi
        \ifnum\outputpenalty>-2000 \else\dosupereject\fi
}
\def\doubleformat{\shipout\vbox{\makeheadline
     \fullline{\box\leftcolumn\hfil\columnbox}\makefootline
     }\advancepageno}
\def\columnbox{\leftline{\pagebody}}
\nopagenumbers
\hfuzz=3pt}
\def\twocolphyzzx{
\def\papersize{\hsize=28pc \fullhsize 9.8in \vsize=6in
   \hoffset=-1 in \voffset=0 in
   \advance\hoffset by .5in \advance\voffset by .5 in
   \skip\footins=\bigskipamount \singlespace }
\Tenpoint 
\let\lr=L
\output={
    \if L\lr
      \global\setbox\leftcolumn=\columnbox
      \global\let\lr=R
    \else
      \doubleformat
      \global\let\lr=L
    \fi
    \ifnum\outputpenalty>-20000
    \else\dosupereject\fi
}
\def\fullline{\hbox to \fullhsize}
\def\doubleformat{\shipout\fullline{\box\leftcolumn\hfil\columnbox}}
\def\columnbox{\vbox{\leftline{\pagebody}\makefootline}\advancepageno}
}
\gdef\twocollarge{
\def\papersize{\hsize=56pc \fullhsize 19.6in \vsize=12in
   \hoffset=-2 in \voffset=0 in
   \advance\hoffset by 1in \advance\voffset by 1 in
   \skip\footins=\bigskipamount \singlespace }
\def\makefootline{\baselineskip=24pt \fullline{\the\footline}}
\let\lr=L
\output{\if L\lr
   \global\setbox\leftcolumn=\columnbox \global\let\lr=R 
  \else \doubleformat \global\let\lr=L\fi
        \ifnum\outputpenalty>-2000 \else\dosupereject\fi
}
\def\doubleformat{\shipout\vbox{\makeheadline
     \fullline{\box\leftcolumn\hfil\columnbox}\makefootline
     }\advancepageno}
\def\columnbox{\leftline{\pagebody}}
\def\makefootline{\bigskip\fullline{\the\footline}}
}
%

%
%
%
%
%

%
%

%
%
\def\IIcollarge{
\newlinechar=`\^^J
\immediate\write16{^^J LARGE TWO COLUMN OUTPUT  ^^J}
\immediate\write16{^^J USE dvips -t landscape -x 500 ^^J}
\def\mycolnumber{2}
\input phyzzx
\catcode`\@=11 
\Mypoint
\twocollarge
\PHYSREV
\paperfootline={\hss\ifp@genum\seventeenrm\folio\hss\fi}
\Mypoint
\def\square{\mathchoice\sqr56\sqr56\sqr{2.1}3\sqr{1.5}3}
\def\vev{vacuum expectation value}
\rightline{UCRHEP-T\ucrnum}
{\titlepage
\vskip -.2 in
\title{ {\bigboldv \thetitle }}
\singlespace
\author{\mycp \theauthor }
\abstract \theabstract
\endpage} \frontpagefalse
}
%
%

%
%
\def\ucrnum{134}
\def\thetitle{\break\break
TWO PHOTON PROCESSES AND EFFECTIVE LAGRANGIANS WITH AN
EXTENDED SCALAR SECTOR}
\def\theabstract{We consider the \sm\ with an extended scalar sector, and
study the possible effects of the physics underlying such a model using
an effective lagrangian parametrization. It is found that certain two
photon processes offer windows where such heavy interactions might
be glimpsed, but the realization of this expectation requires enormous
experimental precision.}
\def\theauthor{{ \singlespace
\author{\cp M.A. P\'erez} \address{Departamento de F\'\i sica\break
Centro de Investigaci\'on y de Estudios Avanzados del IPN \break
Apdo. Postal 14-740, 07000 M\'exico D.F., M\'exico.}
\andauthor{\cp J.J. Toscano} \address{Facultad de Ciencias F\'\i sico
Matem\'aticas \break  Universidad Aut\'onoma de Puebla \break
Apdo. Postal 1152, Puebla, Pue., M\'exico.}
\andauthor{\cp J. Wudka} \UCR}}


\newlinechar=`\^^J
\immediate\write16{^^J ONE COLUMN OUTPUT  ^^J}
\def\mycolnumber{1}
\input phyzzx
\catcode`\@=11 
\Twelvepoint
\PHYSREV
\def\square{\mathchoice\sqr56\sqr56\sqr{2.1}3\sqr{1.5}3}
\def\vev{vacuum expectation value}
\rightline{UCRHEP-T\ucrnum}
{\titlepage
\vskip -.2 in
\title{ {\bigboldiii \thetitle}}
\doublespace
\author{\theauthor}
\abstract
\bigskip
\singlespace
\theabstract
\endpage} 
\PHYSREV
\def\whatjournal{P}
\newlinechar=`\^^J

\def\ordernpb#1#2#3{{\bf#1} (#3) #2}
\if P\whatjournal {\global\def\order#1#2#3{\orderprd{#1}{#2}{#3}}}
                    \immediate\write16{^^J PRD references ^^J}\else
                   {\global\def\order#1#2#3{\ordernpb{#1}{#2}{#3}}}
                    \immediate\write16{^^J NPB references ^^J}
\fi

\def\ijmpa#1#2#3{{\it Int. J. of Mod. Phys. {\bf A}}\order{#1}{#2}{#3}}
\def\nim#1#2#3{{\it Nucl. Instrum. \& Methods {\bf B}}\order{#1}{#2}{#3}}
\def\npb#1#2#3{{\it Nucl. Phys. {\bf B}}\order{#1}{#2}{#3}}

\def\plb#1#2#3{{\it Phys. Lett. {\bf B}}\order{#1}{#2}{#3}}
\def\pl#1#2#3{{\it Phys. Lett.\ }\order{#1}{#2}{#3}}
\def\pr#1#2#3{{\it Phys. Rev.\ }\order{#1}{#2}{#3}}
\def\prep#1#2#3{{\it Phys. Rep.\ }\order{#1}{#2}{#3}}
\def\prl#1#2#3{{\it Phys. Rev. Lett.\ }\order{#1}{#2}{#3}}

\def\prd#1#2#3{{\it Phys. Rev. {\bf D}}\order{#1}{#2}{#3}}

\def\zphys#1#2#3{{\it Z. Phys. {\bf C}}\order{#1}{#2}{#3}}
\REF\basic{
S. Weinberg, {\it Physica} 96{\bf A} (1979) 327.
H. Georgi, \npb{361}{339}{1991}; \ibid{363}{301}{1991}.
J. Polchinski, lectures presented at {\it TASI 92}, Boulder, CO, Jun 3-28,
1992.
}
\REF\lefref{
A. De R\'ujula \etal, \npb{384}{3}{1992}.
M.B. Einhorn and J. Wudka, in {\it Workshop on Electroweak
               Symmetry Breaking}, Hiroshima, Nov. 12-15 (1991);
               in {\it Yale Workshop on Future Colliders,} Oct. 2-3 (1992).
               M.B. Einhorn, in {\it Conference on Unified Symmetry
               in the Small and in the Large,} Coral Gables, Fl, Jan.
               25-27 (1993).
               M.B. Einhorn, plenary talk given at the {\it Workshop on
               Physics and Experimentation with Linear $e^+e^-$ Colliders},
               Waikoloa, Hawaii, April 26-30, 1993. Univ. of Michigan
               report UM-TH-93-17, to be published in the proceedings.
               J. Wudka, in {\it Electroweak Interactions and Unified
               Theories,} XXVIII Recontres de Moriond Les Arcs, Savoie,
               France, March 13-20 (1993).
C.P. Burgess and D. London \prl{69}{3428}{1992}.
A. Dobado \etal, \plb{235}{129}{1990}; \zphys{50}{465}{1991}.
M.J. Herrero, in  {\it 1st Int. Triangle  Workshop:
             Standard Model and Beyond: From LEP to UNK and LHC},
             Dubna, USSR, Oct 1-5, 1990; and references therein.
J. Bagger \etal, \npb{339}{364}{1993}.
K. Hagiwara \etal, \prd{48}{2182}{1993}.}
\REF\review{J. Wudka, \ijmpa9{2301}{1994}}
\REF\chiral{
S. Coleman \etal, \pr{177}{2239}{1969}.
C.G. Callan \etal, \pr{177}{2247}{1969}.
C. Bernard, \prd{23}{425}{1981}.
A. Longhitano, \npb{188}{118}{1981}.
J. Gasser and H. Leutwyler, \npb{250}{465}{1985}.
M. Chanowitz \etal, \prd{36}{1490}{1987}.
M.Golden and L. Randall, \npb{361}{3}{1991}.
Georgi, Ref. \basic.
A. Dobado, M.J. Herrero, J. Bagger \etal, Ref. \lefref. 
R.D. Peccei and X. Zhang, \npb{337}{269}{1990}.
A. Pich, lectures presented at the {\it V Mexican School of
         Particles and Fields}, Guanajuato, M\'exico, Dec. 1992.
T. Appelquist and G.-H. Wu, \prd{48}{3235}{1993}.}
\REF\frere{J.-M. Fr\`ere \etal, \plb{292}{348}{1992}.}
\REF\mssm{H. E. Haber and  G.L. Kane,\prep{117}{75}{1985}.
H.E. Haber and J.F. Gunion, \npb{272}{1}{1986} ({\it erratum}:
\npb{402}{567}{1993}); \npb{278}{449}{1986}.}
\REF\dec{
T. Appelquist and J. Carazzone, \prd{11}{2856}{1975}.
J.C. Collins \etal, \prd{18}{242}{1978}.
For a pedagogical introduction see J.C.  Collins, {\it Renormalization}
          (Cambridge U. Press, Cambridge 1984).}
\REF\discrete{S.L. Glashow and S. Weinberg, \prd{15}{1958}{1977}.}
\REF\bw{
W. B\"uchmuller and D. Wyler, \npb{268}{621}{1986}.
W. B\"uchmuller \etal, \plb{197}{379}{1987}.
C.J.C.  Burges and H.J. Schnitzer, \npb{228}{464}{1983}
C.N. Leung \etal, \zphys{31}{433}{1986}. (1986) 433.}
\REF\hhg{J.F. Gunion \etal, {\it The Higgs Hunter's Guide},
        (Addison-Wesley, Redwood City, CA, 1990)}
\REF\aewlooppap{C. Arzt, \etal, report UM-TH-94-15,
UCRHEP-T125, CALT-68-1932; to appear in Nucl. Phys. B.}
\REF\ren{S. Weinberg, H. Georgi, Ref. \basic.}
\REF\gamgam{
F.R. Arutyunian and V.A. Tumanian, \pl{4}{176}{1963}.
R.H. Milburn, \prl{10}{75}{1963}.
I.F. Ginzburg \etal, \npb{228}{285}{1983}.
I.F. Ginzburg \etal, \nim{205}{47}{1983}; \nim{219}{5}{1984};
          \nim{A294}{72}{1990}.}
\REF\derujula{ A. DeR\'ujula \etal, \npb{384}{3}{1992}.
M. A. P\'erez and J. J. Toscano, \plb{289}{381}{1992}.
K. Hagiwara \etal, \plb{318}{155}{1993}.}
\REF\contreras{J.M. Hernandez, M. A. Perez and
J. J. Toscano, CINVESTAV report FIS-12-94 (unpublished).}
\REF\effeom{
C. Arzt, report UM-TH-92-28, hep-ph/9304230 (unpublished).
C. Grosse Knetter, report BI-TP-93-56, hep-ph/9311259 (unpublished).}
\REF\okun{L.B. Okun, {\sl Leptons and Quarks} (North Holland, Amsterdam
1984) and references therein.}
\def\slacpub{\afterassignment\slacp@b\toks@}
\def\UCRpub{\afterassignment\UCRp@b\toks@}
\def\slacp@b{\Pubnum=\expandafter{UCD--\the\toks@}}
\def\UCRp@b{\Pubnum=\expandafter{UCRHEP-T\the\toks@}}

\def\memohead{\line{\quad\fourteenrm SLAC MEMORANDUM\hfil
       \twelverm\the\date\quad}}
\def\memorule{\par \medskip \hrule height 0.5pt \kern 1.5pt
   \hrule height 0.5pt \medskip}
\def\SLACHEAD{\setbox0=\vtop{\baselineskip=10pt
     \ialign{\eightrm ##\hfil\cr
        \slacbin\cr
        P.~O.~Box 4349\cr
        Stanford, CA 94305\cropen{1\jot}
        (415) 854--3300\cr }}
   \setbox2=\hbox{\caps Stanford Linear Accelerator Center}%
   \vbox to 0pt{\vss\centerline{\seventeenrm STANFORD UNIVERSITY}}
   \vbox{} \medskip
   \line{\hbox to 0.7\hsize{\hss \lower 10pt \box2 \hfill }\hfil
         \hbox to 0.25\hsize{\box0 \hfil }}\medskip }

\FromAddress={\crcr \slacbin \cr
   \P.\ O.\ Box 4349\cr Stanford, California 94305\cr }
\def\slacbin{SLAC\ifx\binno\relax \else , Bin \binno \fi }
\def\binno{81}
\VOFFSET=33pt
\papersize
%
%
\newwrite\figscalewrite
\newif\iffigscaleopen
\newif\ifgrayscale
\newif\ifreadyfile

\def\parsefilename{\ifreadyfile \else
    \iffigscaleopen \else \gl@bal\figscaleopentrue
       \immediate\openout\figscalewrite=\jobname.scalecon \fi
    \toks0={ }\immediate\write\figscalewrite{%
       \the\p@cwd \the\toks0 \the\p@cht \the\toks0 \the\picfilename }%
    \expandafter\p@rse \the\picfilename..\endp@rse \fi }
\def\p@rse#1.#2.#3\endp@rse{%
   \if*#3*\dop@rse #1.1..\else \if.#3\dop@rse #1.1..\else
                                \dop@rse #1.#3\fi \fi
   \expandafter\picfilename\expandafter{\n@xt}}
\def\dop@rse#1.#2..{\count255=#2 \ifnum\count255<1 \count255=1 \fi
   \ifnum\count255<10  \edef\n@xt{#1.PICT00\the\count255}\else
   \ifnum\count255<100 \edef\n@xt{#1.PICT0\the\count255}\else
                       \edef\n@xt{#1.PICT\the\count255}\fi\fi }
\def\redopicturebox{\edef\picturedefinition{\ifgrayscale
     \special{insert(\the\picfilename)}\else
     \special{mergeug(\the\picfilename)}\fi }}
%
%

%



%
%
\font\myrmi=cmr10 scaled\magstep5

\font\mybfi=cmbx10 scaled\magstep5

\font\myii=cmmi10 scaled\magstep5     \skewchar\seventeeni='177
      \skewchar\seventeeni='177
\font\mysyi=cmsy10 scaled\magstep5    \skewchar\seventeensy='60
\font\mysyi=cmsy10 scaled\magstep4    \skewchar\seventeensy='60
\font\myexi=cmex10 scaled\magstep5

\font\mysli=cmsl10 scaled\magstep5

\font\myiti=cmti10 scaled\magstep5

\font\mytti=cmtt10 scaled\magstep4
\font\mytt=cmtt10 scaled\magstep3
\font\mycpi=cmcsc10 scaled\magstep5
\font\mycp=cmcsc10 scaled\magstep4
%
%
\def\seventeenf@nts{\relax
    \textfont0=\seventeenrm          \scriptfont0=\twelverm
      \scriptscriptfont0=\ninerm
    \textfont1=\seventeeni           \scriptfont1=\twelvei
      \scriptscriptfont1=\ninei
    \textfont2=\seventeensy          \scriptfont2=\twelvesy
      \scriptscriptfont2=\ninesy
    \textfont3=\seventeenex          \scriptfont3=\fourteenex
      \scriptscriptfont3=\twelveex
    \textfont\itfam=\seventeenit     \scriptfont\itfam=\twelveit
    \textfont\slfam=\seventeensl     \scriptfont\slfam=\twelvesl
    \textfont\bffam=\seventeenbf     \scriptfont\bffam=\twelvebf
      \scriptscriptfont\bffam=\ninebf
    \textfont\ttfam=\mytt
    \textfont\cpfam=\mycp }
\def\seventeenpoint{\seventeenf@nts \samef@nt \b@gheight=17pt \setstr@t }
\def\myf@nts{\relax
    \textfont0=\myrmi          \scriptfont0=\seventeenrm
      \scriptscriptfont0=\twelverm
    \textfont1=\myii           \scriptfont1=\seventeeni
      \scriptscriptfont1=\twelvei
    \textfont2=\mysyi          \scriptfont2=\seventeensy
      \scriptscriptfont2=\twelvesy
    \textfont3=\myexi          \scriptfont3=\seventeenex
      \scriptscriptfont3=\twelveex
    \textfont\itfam=\myiti     \scriptfont\itfam=\seventeenit
    \textfont\slfam=\mysli     \scriptfont\slfam=\seventeensl
    \textfont\bffam=\mybfi     \scriptfont\bffam=\seventeenbf
      \scriptscriptfont\bffam=\twelvebf
    \textfont\ttfam=\mytti
    \textfont\cpfam=\mycpi
\chapterfontstyle={\mybfi}
\def\smallii{\smallv}
}
\def\mypoint{\myf@nts \samef@nt \b@gheight=18pt \setstr@t }
\newif\ifmy@
\def\Mypoint{\mypoint\my@true\twelv@false\spaces@t}
\def\Tenpoint{\tenpoint\my@false\twelv@false\spaces@t}
\def\Twelvepoint{\twelvepoint\my@false\twelv@true\spaces@t}
\def\spaces@t{\rel@x
\ifmy@
    \ifsingl@\subspaces@t8:7;\else\subspaces@t9:5;\fi
  \else
\iftwelv@
    \ifsingl@\subspaces@t3:4;\else\subspaces@t1:1;\fi
  \else
    \ifsingl@\subspaces@t3:5;\else\subspaces@t4:5;\fi
\fi \fi
\ifdoubl@ \multiply\baselineskip by 5 \divide\baselineskip by 4 \fi
}
\def\normalbaselines{ \baselineskip=\normalbaselineskip
   \lineskip=\normallineskip \lineskiplimit=\normallineskip
   \iftwelv@ \else
   \ifmy@
\multiply\baselineskip by 5 \divide\baselineskip by 4
     \multiply\lineskiplimit by 5 \divide\lineskiplimit by 4
     \multiply\lineskip by 5 \divide\lineskip by 4
\else
\multiply\baselineskip by 4 \divide\baselineskip by 5
     \multiply\lineskiplimit by 4 \divide\lineskiplimit by 5
     \multiply\lineskip by 4 \divide\lineskip by 5 \fi \fi}
\def\titlestyle#1{\par\begingroup \titleparagraphs
\ifmy@\mypoint\else\iftwelv@\fourteenpoint\else\twelvepoint\fi\fi
\noindent #1\par\endgroup }
\def\refout{\par\penalty-400\vskip\chapterskip
   \spacecheck\referenceminspace
   \ifreferenceopen \Closeout\referencewrite \referenceopenfalse \fi
   \line{\ifmy@\myrmi\else\iftwelv@\fourteenrm\else\twelverm\fi\fi
   \hfil REFERENCES\hfil}\vskip\headskip
   \input \jobname.refs
   }
\def\figout{\par\penalty-400
   \vskip\chapterskip\spacecheck\referenceminspace
   \iffigureopen \Closeout\figurewrite \figureopenfalse \fi
   \line{\ifmy@\myrmi\else\iftwelv@\fourteenrm\else\twelverm\fi\fi
   \hfil FIGURE CAPTIONS \hfil}\vskip\headskip
   \input \jobname.figs
   }
\def\tabout{\par\penalty-400
   \vskip\chapterskip\spacecheck\referenceminspace
   \iftableopen \Closeout\tablewrite \tableopenfalse \fi
   \line{\ifmy@\myrmi\else\iftwelv@\fourteenrm\else\twelverm\fi\fi
   \hfil TABLE CAPTIONS \hfil}\vskip\headskip
   \input \jobname.tabs
   }
\def\address#1{\par\kern 5pt\titlestyle{\it #1}}
  \def\Vfootnote#1{%
      \insert\footins%
      \bgroup%
         \interlinepenalty=\interfootnotelinepenalty%
         \floatingpenalty=20000%
         \singl@true\doubl@false%
         \ifmy@\seventeenpoint\spaces@t
         \smallskip
         \else\Tenpoint\fi%
         \splittopskip=\ht\strutbox%
         \boxmaxdepth=\dp\strutbox%
         \leftskip=\footindent%
         \rightskip=\z@skip%
         \parindent=0.5%
         \footindent%
         \parfillskip=0pt plus 1fil%
         \spaceskip=\z@skip%
         \xspaceskip=\z@skip%
         \footnotespecial%
         \Textindent{#1}%
         \footstrut%
         \futurelet\next\fo@t%
   }
\paperfootline={\hss\ifmy@\seventeenrm\folio\hss%
\else\iffrontpage\else\ifp@genum\tenrm\folio\hss\fi\fi\fi}

\noblackbox
\doublespace
\def\thecaption#1#2{\centerline{\vbox to 1 in{\hsize 5 in \vfill
           {\ifmy@\seventeenpoint\spaces@t\else\Tenpoint\fi
           \textindent{#1} \global \advance \baselineskip by -12 pt
            \noindent#2 }}}\global \advance \baselineskip by 12 pt}
\def\effl{{effective lagrangian}}
\def\al{y}

\def\ao{{a_0}}
\def\aogg{\ao\hbox{-}\gamma\hbox{-}\gamma }
\def\mao{m_\ao}
\def\mw{m\lowti{w}}
\def\mh{m_{H^+}}
\def\mf{m_f}
\def\me{m\lowti{e}}
\def\mt{m\lowti{t}}
\def\mb{m\lowti{b}}
\def\oti{\ocal\lowti{tree}{}_1}
\def\otii{\ocal\lowti{tree}{}_2}
\def\ppi{\pcal_1}
\def\ppii{\pcal_2}
\def\sii{\sqrt{2}}
\def\hho{{H_0}}
\def\ho{{h_0}}
\def\hp{H^+}

\def\impp{\im\left(\phi^\dagger_1 \phi_2\right)}
\def\gp{G^+}
\def\gm{G^-}
\def\go{G_0}
\def\ca{c_\alpha}
\def\sa{s_\alpha}
\def\cb{c_\beta}
\def\sb{s_\beta}
\def\sw{s\lowti{w}}
\def\cw{c\lowti{w}}
\def\al{y}
\def\camb{ c_{ \alpha - \beta} }
\def\capb{ c_{ \alpha + \beta} }
\def\captb{ c_{ \alpha + 3 \beta } }
\def\samb{ s_{ \alpha - \beta} }
\def\sapb{ s_{ \alpha + \beta} }
\def\saptb{ s_{ \alpha + 3 \beta } }
\def\ctb{ c_{ 2 \beta } }
\def\qwe{ \eta }
\catcode`\@=12 

\chapter{Introduction}

The use of effective lagrangians~\refmark\basic\
in parametrizing physics beyond a given
theory has been studied extensively in the recent literature, especially
within the context of the \sm.~\refmark\lefref\
The formalism generates a model-independent parametrization
of the low energy limit of any Green function in terms of a
set of unknown constants, each of which multiplies a local operator
involving only the low energy fields; all symmetries are also naturally
preserved.
Despite the fact that the formalism involves, in principle, an infinite
number of local operators in the light fields, only a finite
number of them need to be considered in any given calculation;
the number of operators which are considered is determined
by the required degree of accuracy: for higher precision
more terms in the effective lagrangian must be included, and the number
of parameters increases. Any two models of describing the heavy physics
will generate low energy Green functions whose expressions fit
the effective lagrangian parametrization (the explicit values of
the parameters are, of course, model dependent).

The requirement that the low energy particle content should be the same
as in the \sm\ is clearly an assumption. One can envisage, for example,
the case
where the Higgs particle is either heavy or absent [\chiral], or a
situation where there is an additional relatively light vector boson
[\frere]. In this paper we will consider a different modification of
the low energy lagrangian for which the low energy
scalar sector contains two
doublets. Such a model, motivated by various supersymmetric extensions
of the \sm\ [\mssm], will be our
low energy theory; the adjective ``light'' will refer to it.

Starting from this low energy theory
we will investigate the possible effects of (heavy) physics underlying
it. The purpose of this investigation is dual:
on the one hand
we will determine the difficulties in uncovering heavy physics effects
even in some very favorable circumstances, which will be of use when and if
the minimal supersymmetric extension of the \sm\ is verified
experimentally. On the other hand the calculations presented in this paper
provide a good illustration of the uses of effective operators within
the context of renormalization theory.

Since the light theory is renormalizable and has light scalar
excitations (whose masses are not protected against $O ( \Lambda )$
radiative corrections), the underlying physics is
expected to be light and decoupling [\review]. We will denote by $
\Lambda $ the scale at which the underlying physics becomes apparent,
then all observables can be expanded in a power series in $ 1 / \Lambda
$ [\dec].~\foot{ Note that in the process of
integrating out the heavy physics some quantities will get corrections
which do not vanish as $ \Lambda \rightarrow \infty $, but all these
contributions can be absorbed in a renormalization of the two doublet
model parameters and are therefore
unobservable. The effects of these contributions
affect the naturality of the model and will no be considered further in this
paper.}

The two doublet model is known to have problems with flavor changing
neutral currents. We chose to alleviate these difficulties by imposing
the usual discrete symmetry [\discrete]. We will for simplicity also
impose this symmetry on the
heavy operators (within the framework of this paper this means that the
processes  which violate this symmetry occur at a scale $ \Lambda'
\gg \Lambda $, which is supported by explicit computation of the bounds
on the scale of process which mediate flavor changing neutral currents
[\bw]). We will argue below that the same estimates of the sensitivity
of a given experiment to $ \Lambda $
are obtained even
when the underlying physics does not obey the discrete symmetry. We will
also ignore CP violation in the light theory; more precisely, we will
assume that the scale of CP violation is $ \gesim \Lambda $, as
discussed in section 4.

A general study of the two doublet model together with all operators
of dimension $ > 4 $ would be both prohibitive and obscure. Fortunately
this is not necessary: most processes receive light physics contributions
(\ie, those generated by the two-doublet extension of the \sm)
which dominate any contribution from the heavy physics.
Thus we need only
concentrate on those processes which are suppressed
within the two-doublet model and which acquire significant contributions
due to the inclusion of effective operators.

As an example of the above situation we will
study the two photon decay  of the CP-odd scalar $ \ao $ present
in this theory, as well as $ \ao $ production in polarized
photon colliders. Since this particle has no couplings to the
gauge bosons within the light theory, the usual contribution to these process
are mediated
by fermion loops. Explicit calculation as well as simple arguments
show that these graphs are sizable only for the top quark contribution
which is proportional to~\foot{ $ \tan \beta $ denotes
the ratio of the \vev s, see below.} $ \cot \beta $ and
will be small in the $ \beta \rightarrow \pi/2 $ limit.
Alternatively we will see that the production of the $ \ao $
within the light theory in photon collisions (via quark loops)
is suppressed for a certain
combination of photon polarizations. This, however, is not the case for
the amplitude induced by the effective operators. We can then
arrange for a suppression of the light-theory production amplitude,
with the corresponding enhancement of the ``anomalous'' signal,
by choosing the photon polarizations appropriately.~\foot{The constants
multiplying the effective operators are estimated using standard
arguments (see below); it is of course possible for some of these
constants to be suppressed due to unknown effects, we will comment on
this possibility later on.}

The organization of the paper is as follows. In the next section
we will construct the relevant operators for the above processes.
The calculation of the corresponding amplitudes together with some
illustrative numerical results is presented in
section 3. Section 4 contains some parting comments. Details of the
calculations are summarized in the appendix.

\chapter{{\caps The Lagrangian}}

We begin with a brief description of the model starting with the
renormalizable interactions; later we turn to the
study of heavy physics effects using effective operators.

The scalar sector contains two doublets $ \phi_a  \ (a=1,2) $, while the
fermionic and gauge particle contents are identical to those
in the \sm. As mentioned in the introduction
we will impose a $ \ZZ_2 $ symmetry on the low energy
sector in order to suppress flavor
changing neutral currents. Under this symmetry $ \phi_1 $
and all right handed fermions are odd, while the remaining
particles are even. We will use the conventions of Refs. \bw\ and \hhg.

The light scalar potential (that is,
the terms in the potential of dimension $ \le 4 $) is given by
$$ \eqalign{ V =& \lambda_1 \left( \phi_1^\dagger \phi_1 -v_1^2 \right)^2
                + \lambda_2 \left( \phi_2^\dagger \phi_2 -v_2^2 \right)^2 \cr
& + \lambda_3 \left( \phi_1^\dagger \phi_1 + \phi_2^\dagger \phi_2 -
v^2 \right)^2 + \lambda_4 \left[ \left( \phi_1^\dagger \phi_1 \right)
\left( \phi_2^\dagger \phi_2 \right) - \left| \phi_1^\dagger
\phi_2 \right|^2 \right] \cr
& + \lambda_5 \left[ \re \left( \phi_1^\dagger \phi_2 \right) -
v_1 v_2 \right]^2 + \lambda_6 \left[ \im \left( \phi_1^\dagger \phi_2 \right)
\right]^2 \cr } \eqn\eq $$ In terms of the mass eigenstates the scalar
fields are
$$ \eqalign{
\phi_1^0 &= v_1 + \inv{ \sii } ( \ca \hho - \sa \ho )
        + { i \over \sii } ( \cb \go - \sb \ao )  \cr
\phi_2^0 &= v_2 + \inv{ \sii } ( \sa \hho + \ca \ho )
        + { i \over \sii } ( \sb \go + \cb \ao ) \cr
\phi_1^+ &= \cb \gp - \sb \hp \cr
\phi_2^+ &= \sb \gp + \cb \hp \cr } \eqn\eq $$
where
$$ \phi_a = \pmatrix{ \phi_a^+ \cr \phi_a^0 \cr } , \quad a = 1 , 2 \eqn\eq $$
We used the usual definitions $ \tan \beta = v_2 / v_1 $,
$ v^2 = v_1^2 + v_2^2 $; $ \ca = \cos\alpha $, etc.
The fields $ \go , \ \gp $ and $ \gm $ correspond to the would-be
Goldstone bosons (with $ G^- = \left( G^+ \right)^\dagger $),
while the other scalar fields are physical.

In the calculations below we will adopt the following
notation and conventions (see Ref. \bw): $ \ell $ and $ q $
denote the left-handed lepton and quark doublet respectively, $e , \ u$ and
$d$ denote the corresponding right-handed $ \su2$ singlets,
(of particular interest to
us will be the
case where $u$ denotes the right-handed top quark and $q$ the
left-handed top-bottom doublet). The $ \su
2$ gauge fields are denoted by $ W_\mu^I$ (where $I$ is a weak isospin
index) and the $ \ui $ gauge field is labelled $ B_\mu $, the
corresponding field strengths are~\foot{ $
W_{ \mu \nu }^I $ is the full non-Abelian expression for the curvature.
} $ W_{ \mu \nu }^I $ and $ B_{ \mu \nu } $ ; finally $ \tau^I$ denote the
Pauli matrices
and $ \tilde \phi_k = i \tau^2 \phi_k^* $.
The covariant derivatives are $ D_\mu = \partial_\mu - { i g
\over 2 }  \tau^I W_\mu^I - { i g' \over 2 } Y B_\mu $ where $Y$ is the
hypercharge and $ g ,  \ g' $ the gauge couplings corresponding to
$ \su 2 _L $ and $ \ui_Y $ respectively.

With these preliminaries we can now
discuss the possible effects of
heavy physics on low energy observables (where ``low energy'' refers to
scales below that of symmetry breaking in the above two-doublet model),
concentrating on the corresponding
modifications to the coupling of the CP odd scalar $ \ao $ to
two photons. We will assume that the underlying
physics is decoupling and weakly coupled, and
denote by $ \Lambda $ its characteristic
scale (the scale at which it becomes directly observable).
We only consider the weakly coupled case since for strongly
coupled underlying physics
it is difficult to maintain the scalars $ \phi_a $ naturally light
(compared to  $ \Lambda $) without considerably
modifying the low energy spectrum [\review].
The decoupling theorem
[\dec] then implies that all the effects of heavy
excitations at low energies can be parametrized
in a model independent manner by a series of local operators
which satisfy the same symmetries as the low energy theory,
and whose coefficients are suppressed by the appropriate power of
$ \Lambda $. It is of course possible for the underlying theory to have
several mass scales, if this is the case $ \Lambda $ will denote the
smallest of these scales.~\foot{ This implies that a suppression
factor $ \sim ( \Lambda / \Lambda' )^2 $ may be present in some of the
operators.}

The construction of the \effl\
closely parallels the one described in Ref.\bw.
The terms of dimension four are the ones in the two-doublet
model; there are no terms of dimension five due to Lorentz and gauge
symmetries (right-handed neutrinos are assumed to be
absent). Therefore all the effects of the heavy excitations will be
suppressed by at least two powers of $ \Lambda $.  The
simplest way of obtaining these operators is to take the list provided
in Ref.\bw\ and replace the \sm\ doublet by either $
\phi_1 $ or $ \phi_2 $~\foot{We have verified that this procedure generates
all the operators relevant for
the calculations below, despite the dfact that the equations of motion
were used in Ref. \bw\ to eliminate some operators.}.

Each of the effective operators appears in the \effl\ multiplied by
an undetermined coefficient. This apparently implies that there
is no way of extracting from this  formalism quantitative estimates.
Fortunately this is not the case: having taken the underlying physics
to be weakly coupled, the order of magnitude of the
coefficient of an operator is determined by the type
of graph in the underlying theory which generates it. If an operator
is generated at tree level by the heavy dynamics, it will appear with
a coefficient $ \sim h / \Lambda^2 $ where $h$
denotes a product of coupling constants in the underlying theory; these
by assumption are $ \lesim 1 $, so that we also expect $ h \lesim 1 $.
Loop-generated operators
will acquire an extra suppression factor, so that their coefficient
are of the form $ h' / ( 4 \pi \Lambda )^2 $ with $h' \lesim 1 $.
The additional loop factor
$ \sim 1/ ( 4 \pi)^2 $ insures that the loop-generated operators
are subdominant. We will
assume that the heavy physics is described by a gauge theory
and, for quantitative estimates will take the corresponding
gauge coupling to be $ \sim 1 $; one should keep in mind that
the results thus obtained represent an upper bound and could
be suppressed by a significant amount.

The standard (two-doublet) contribution to the $ \aogg $
coupling occurs at one loop [\hhg]. The only dimension six
operators which generate tree-level contributions to this vertex
are
$$ \eqalign{
&\hfil \oti =  \left(  B_{ \mu \nu } \right)^2 \; \impp , \hfil \qquad
\tilde \oti =B_{ \mu \nu } \tilde B_{ \mu \nu } \; \impp , \hfil \cr
& \hfil \otii =  \left( W^I _{ \mu \nu } \right)^2 \; \impp , \hfil \qquad
\tilde \otii =   W^I_{ \mu \nu } \tilde W^I_{ \mu \nu } \; \impp ; \hfil \cr }
\eqn\otree$$
(together with their hermitian-conjugate counterparts), which are
loop-generated [\aewlooppap]: there are no dimension six,
tree-level-generated operators that contribute to the $ \aogg$ vertex at tree
level.
Note that these operators violate the $\ZZ_2 $ symmetry.
A consistent loop expansion then requires that we include
all  tree-level contributions containing one \otree\ insertion.
We must also include all one-loop graphs containing
one $ \ocal$ insertion, provided this operator is generated at
tree level. The one loop graphs containing insertions of loop-generated
operators can be ignored; this considerably reduces the number
of terms that need to be considered.

At this point we can either assume that the underlying dynamics respects
the discrete symmetry or not. The precise results differ depending on
which of these cases is realized; note however that in either case
all contributions to the vertex of interest occur at one loop; either
through tree-level-generated operators in one loop graphs, or through
loop-generated operators in tree level graphs.
In this paper we are only
interested in estimating the effects of the underlying
physics and for this it is sufficient, in view of the previous comments,
to consider the situation where the underlying physics
obeys the discrete symmetry. This assumption has the added
advantage of simplifying the calculations.

When the physics underlying the two
doublet model is assumed to respect the discrete symmetry there
are no tree-level contributions to the $ \aogg $  vertex since
the operators in \otree\ are forbidden. It follows that
the dimension six modifications of the low energy lagrangian
contribute to the processes under consideration only
through (explicit) one loop graphs; as mentioned
above only tree-level-generated operators need be considered.

Due to the
absence of tree level interactions, the loop contributions
must be {\it finite}. Indeed, any divergence can be
associated with  a local operator respecting the symmetries
of the model; the effective lagrangian already contains
all such operators, and all divergences can be absorbed in a renormalization of
the effective lagrangian coefficients [\ren]. Since there is no tree level
operator (when the $ \ZZ_2 $ symmetry is imposed) containing an $ \aogg $ term,
it follows that the corresponding three point function must
be finite, else it would be impossible to absorb its divergence.
The cancellation of divergences, together with electromagnetic gauge
invariance, provide important checks on the algebra. Imposing the
discrete symmetry on the effective operators also implies that
there is no dependence (to one loop) on the renormalization scale.

The effective lagrangian takes the form
$$ \lcal = \lcal \lowti{ two \
doublets } + \inv{ \Lambda^2 } \sum_i \left[
\al_i \ocal_i + \hbox{ h. c. } \right] + O ( 1 / \Lambda^3 )
\eqn\leff $$ where the summation over $i$ runs
over the set of (tree-level-generated)
operators given in the appendix. The unknown constants
$ y_i $ depend on the underlying physics and, based on the
discussion above, are expected to be $ \sim 1 $.

The most important terms in \leff\ are generated by the operators $$
\ocal_{ u \varphi ; 211 } = \left( \phi_2^\dagger \phi_1 \right)
\left( \bar q u \tilde \phi_1 \right),
\quad \ocal_{ i j k l } \up 1 =
               \left[ \partial_\mu \left( \phi^\dagger_i \phi_j
\right) \right] \left[ \phi_k^\dagger { \buildrel \leftrightarrow
\over D_\mu } \phi_l \right] ; \eqn\main $$
which dominate the heavy physics contributions to the
$ \aogg $ vertex (in $ \ocal\up1, \ i,j,k,l=1,2$ and only those terms
with an even number of $ \phi_1 $ factors are considered due to the
discrete symmetry).  The corresponding
contributions are not suppressed in the limit of large $ \tan \beta $;
nor are they suppressed by (explicit) small fermions masses.~\foot{
It is of course possible for the underlying dynamics to generate
coefficients which are small for large $ \tan \beta $
and/or contain small Yukawa couplings, we will not consider this
possibility here.}.

\chapter{Calculation of the amplitudes}

Having described the lagrangian we can use it to evaluate the
three point function for two photons and one $ \ao $.
The photons will be labelled by the subscripts $1$ and $2$,
the corresponding momenta and polarization vectors will be denoted by $ k_a $
and
$ e_a $ ($ a = 1 , 2 $) respectively.
We will assume that all (external) particles are on shell.

Due to electromagnetic gauge invariance there are two possible
tensorial structures allowed for the amplitude:
$$ \ppi = ( k_1 \cdot k_2 ) (e_1 \cdot e_2 ) - ( k_1 \cdot e_2 )
(k_2 \cdot e_1 ) , \qquad
\ppii = i \epsilon_{ \alpha \beta \mu \nu } e_1^\alpha e_2^\beta
k_1^\mu k_2^\nu \eqn\eq $$ correspondingly the amplitude can be
written as $$ \acal_\aogg = \mcal_1 \ppi + \mcal_2 \ppii .\eqn\eq $$

The light-physics amplitude generated
via quark loops contains only
$ \ppii $. In contrast the $ \ocal $ induced terms can be proportional to
either quantity, in particular all bosonic loops give results proportional
to $ \ppi $. This implies that the decay $ \ao \rightarrow \gamma \gamma $
will have a modified width and angular distribution due to the presence
of the operators $ \ocal $. Unfortunately the branching ratio is so small that
the
modification of the angular distribution is unobservable
(at least in the forseeable future).

In contrast, the presence of terms proportional to  $ \ppi $
in the amplitude can have important effects in $ \ao $ production
in photon colliders. These machines will (probably) allow the possibility
of polarizing the photons whence the terms $ \propto \ppii $ can be
suppressed by the appropriate choice of initial polarizations
thus enhancing the heavy physics contributions.

The calculation of the amplitude involves both one particle
irreducible and one particle reducible diagrams. The latter arise due to
the $ \ao - h_0 $ and $ \ao - H_0 $ mixings induced by operators
such as the second one in \main. The main contributions to the 1PI
graphs come from the first operator in \main, which also becomes the overall
dominant term when $ \alpha $ is close to zero or to $ \pi/2 $ (provided
$ \tan \beta $ is large). For intermediate values of $ \alpha $ the
contributions from both operators in \main\ are comparable.~\foot{The
1PR graphs will in general exhibit resonances at the mass of the
$ h_0 $ and $ H_0 $. The previous comments correspond to energies outside
these regions.}

The total amplitude derived from \leff\ is
$$ \acal_{ \ao \rightarrow \gamma \gamma }
= \acal \lowti{light} + \inv{ \Lambda^2 }
\sum_i \al_i \acal_i \eqn\eq $$ where
$ \acal_i $ denotes the contribution to  the $ \aogg $ (on-shell) three point
function generated by $ \ocal_i $ (without the factor
$ \al_i / \Lambda^2 $). From $ \acal_{ \ao \rightarrow \gamma \gamma }
$ the two-photon decay width of the $ \ao $ or its production cross
section in photon collisions
can be evaluated directly. Note that the terms containing $ \ppi $
and $ \ppii $ will not interfere in polarization-averaged widths or
cross-sections. The quantities $ \acal_i $ are displayed in the
appendix.

\FIG\width{
Width and branching ratios
for the decay $ \ao \rightarrow \gamma \gamma $
as a function of $ \mao $ for two values of $ \beta $. The solid curves
correspond to the theory with effective operators for $ \Lambda = 1 \tev $,
the dashed curves give
the pure two-doublet model results. We chose $ m_\ho = 450 \gev $, $
m_\hho = 550 \gev $, $ \mh = 500 \gev $;
$ m \lowti{top}=174 \gev $, $ \alpha=45^o$. The
resonance peaks are produced by the 1PR diagrams.}

\FIG\widthratio{ {\bf(a),(b)}: Ratio of the total $ \ao \rightarrow \gamma
\gamma $ decay width
to the decay width of a \sm\ Higgs particle of the same mass into two
photons. {\bf(c),(d)} Total number of yearly events at the LHC (C.M.
energy$=7$TeV,
luminosity $100$/fb).
The solid curves
correspond to the theory with effective operators for $ \Lambda = 1 \tev $,
the dashed curves give
the pure two-doublet model results. We chose $ m_\ho = 450 \gev $, $
m_\hho = 550 \gev $, $ \mh = 500 \gev $ and $ \Lambda = 1 \tev $;
$ m \lowti{top}=174 \gev $, $ \alpha=45^o$.}

\FIG\lamdep{ Branching ratio for the decay $ \aogg $ for $ \Lambda = 1
\tev $ (solid curve), $ \Lambda = 2 \tev $ (dashed curve) and $ \Lambda =
\infty $ (dotted curve). We chose $ m_\ho = 450 \gev $, $ m_\hho = 550
\gev $, $ \mh = 500 \gev $ and $ \tan\beta = 30 $;
$ m \lowti{top}=174 \gev $, $ \alpha=45^o$. }

\FIG\csect{Total $ \ao $ production cross section averaged over an
energy bin equal to the total width of the $ \ao $. The solid curves
correspond to the theory with effective operators for $ \Lambda = 1 \tev $,
the dashed curves give
the pure two-doublet model results.
 We chose $ m_\ho = 450 \gev $, $
m_\hho = 550 \gev $, $ \mh = 500 \gev $;
$ m \lowti{top}=174 \gev $, $ \alpha=45^o$. The
electron polarization is 70\%.}

\FIG\ratio{ Ratio of the total to light cross sections. The solid curve
correspond to $ \tan \beta = 30$, the dashed curve to $ \tan\beta = 10 $.
We chose $ m_\ho = 450 \gev $, $
m_\hho = 550 \gev $, $ \mh = 500 \gev $ and $ \Lambda = 1 \tev $;
$ m \lowti{top}=174 \gev $, $ \alpha=45^o$. The
electron polarization is 70\% .}

\FIG\eventnumber{Number of $ \ao $ events produced in
$ \gamma \gamma $ colissions for two values of $ \tan\beta $.
We chose $ m_\ho = 450 \gev $, $
m_\hho = 550 \gev $, $ \mao = \mh = 500 \gev $;
$ m \lowti{top}=174 \gev $, $ \alpha=45^o$. The
electron polarization is 70\%  and the luminosity equals 10/fb.}

\FIG\statsign{Statistical significance of the deviations
from the light cross section,
solid curve $ \tan \beta = 30 $, dashed curve
$ \tan\beta = 10 $. We chose $ m_\ho = 450 \gev $, $
m_\hho = 550 \gev $, $ \mao = \mh = 500 \gev $;
$ m \lowti{top}=174 \gev $, $ \alpha=45^o$. The
electron polarization is 70\%  and the luminosity equals 10/fb.}

\expel
\setbox2=\vbox to 2 truein{\epsfxsize=4 in\epsfbox[0 0 612 792]{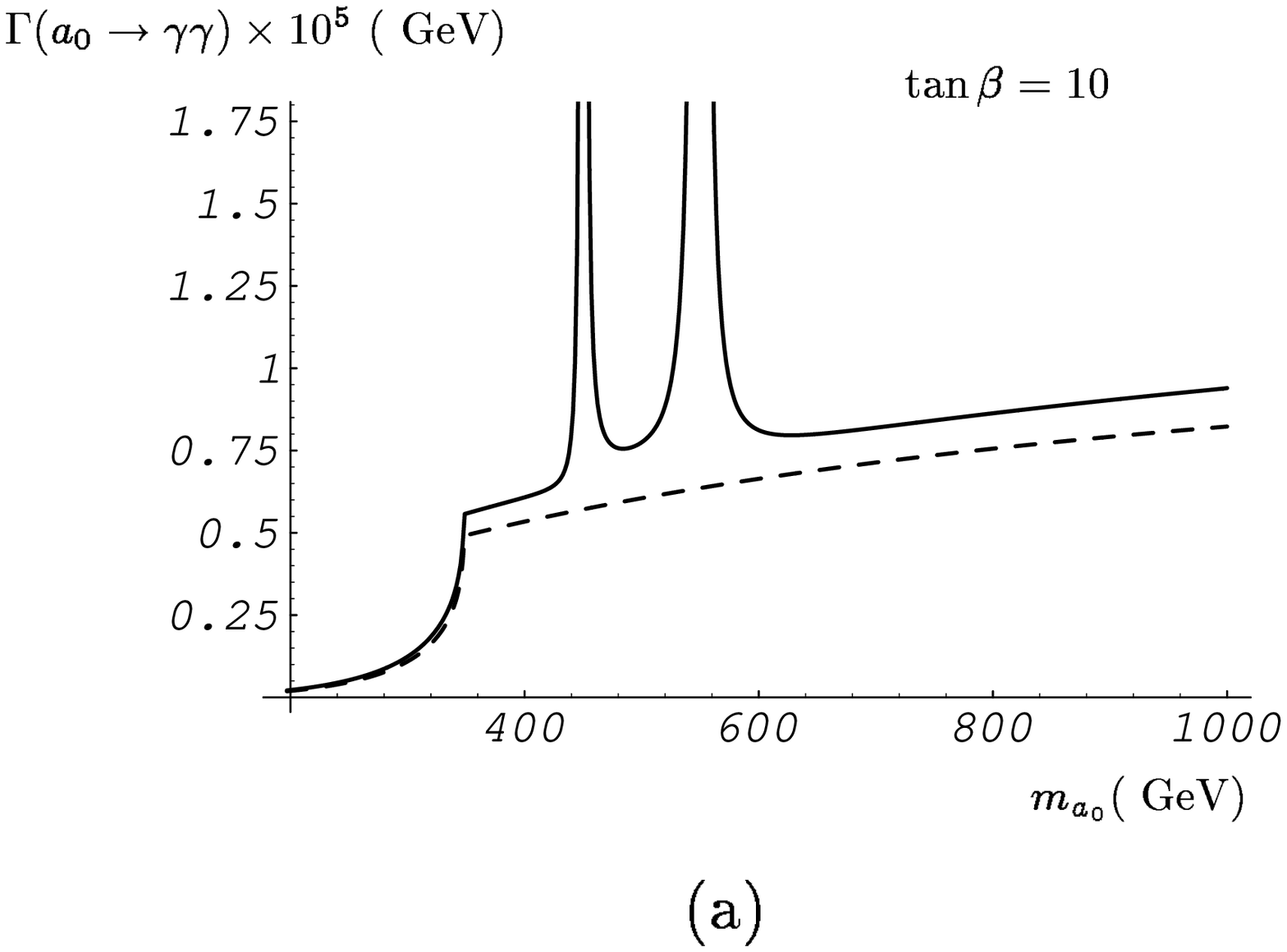}}
\setbox3=\vbox to 2 truein{\epsfxsize=4 in\epsfbox[0 0 612 792]{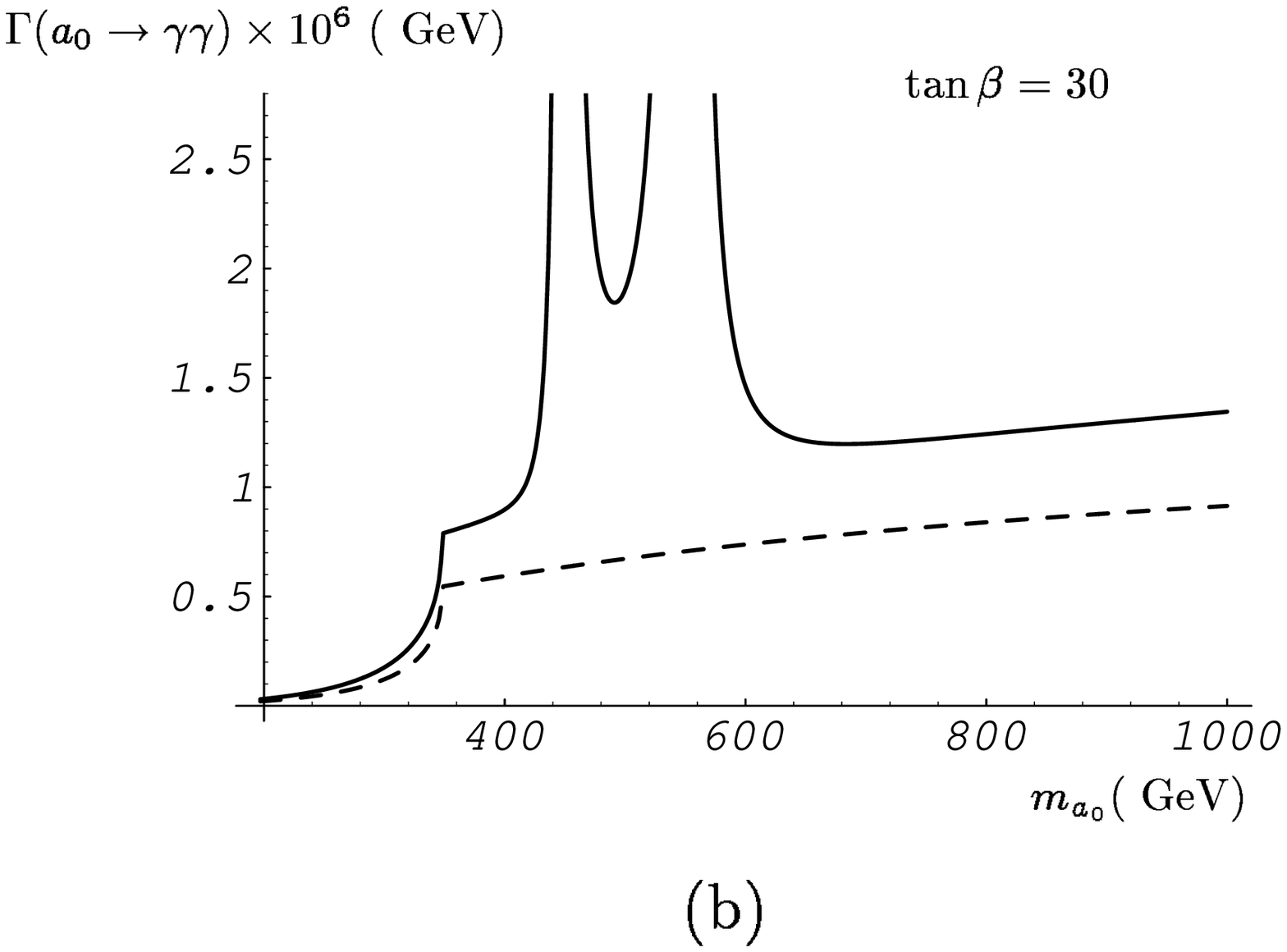}}
\setbox4=\vbox to 2 truein{\epsfxsize=4 in\epsfbox[0 0 612 792]{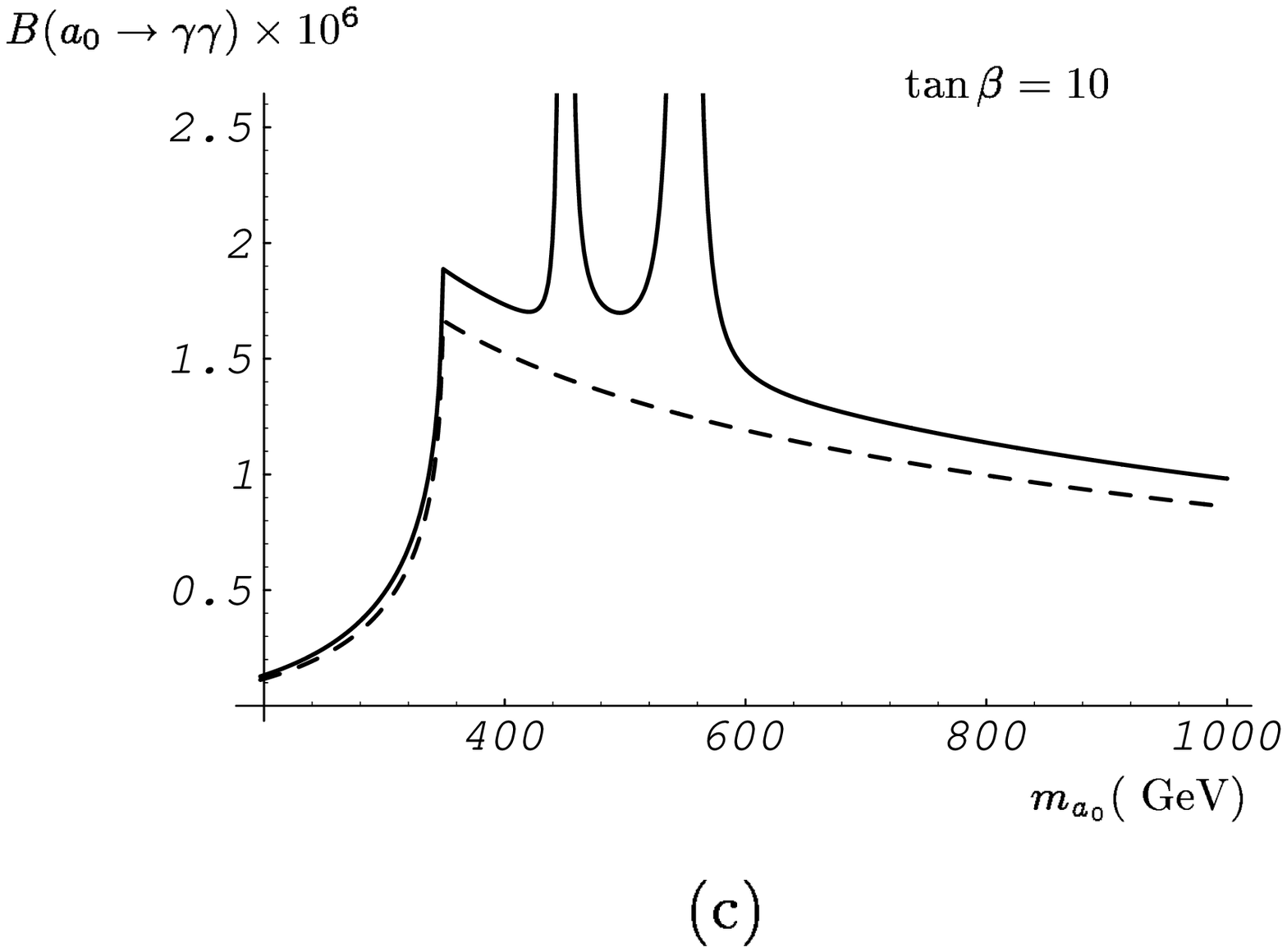}}
\setbox5=\vbox to 2 truein{\epsfxsize=4 in\epsfbox[0 0 612 792]{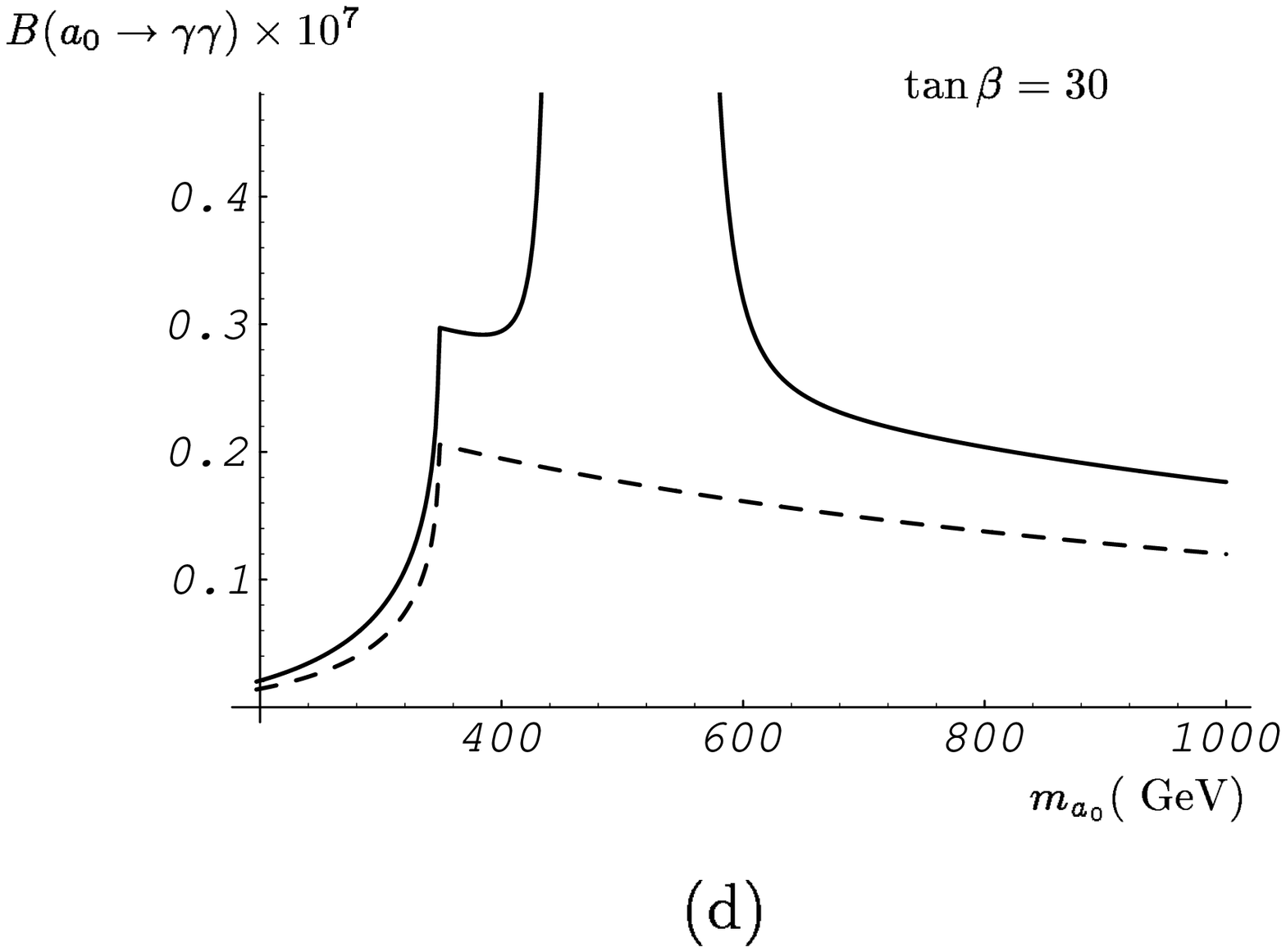}}

\vskip -20 pt

\line{$\mskip-80mu$ \box2 $\mskip-40mu$ \box3 }


\bigskip

\line{$\mskip-80mu$ \box4 $\mskip-40mu$ \box5 }

\bigskip\bigskip\bigskip

\thecaption{Figure 1.}{
Width and branching ratios
for the decay $ \ao \rightarrow \gamma \gamma $
as a function of $ \mao $ for two values of $ \beta $. The solid curves
correspond to the theory with effective operators for $ \Lambda = 1 \tev $,
the dashed curves give
the pure two-doublet model results. We chose $ m_\ho = 450 \gev $, $
m_\hho = 550 \gev $, $ \mh = 500 \gev $;
$ m \lowti{top}=174 \gev $, $ \alpha=45^o$. The
resonance peaks are produced by the 1PR diagrams.}

\bigskip

For generic values of the parameters in the standard two-doublet
lagrangian the contributions from the $ \ocal_i $ are small due
to the presence of the $ 1 / \Lambda^2 $ factor. This suppression
can be countered in two ways. For $ \ao $ production we can use
the photon polarization to suppress the light contribution
and bring the anomalous effects to the foreground. For
the two-photon decay of the $ \ao $ the light contributions
are proportional to either $ \mb \tan \beta $
or $ \mt \cot \beta $ and are suppressed
when $ | \beta - \pi / 2 | $ is similar to zero (though larger
than $ \sim \mb^2 / \mt ^2 $). The only terms which can compete with the
light contributions are those which do not vanish as $ \beta \rightarrow
\pi / 2 $  and are not suppressed by a small mass factor.
These constraints are satisfied only by the operator \main.

Using the results of the appendix we evaluated the $ \ao \rightarrow
\gamma \gamma $ width and branching ratio for several illustrative
choices of the masses and couplings, the results are presented in
Fig. \width~\foot{These results are presented for the choice
of constants $ \al_i $ specified in the Appendix.}.

\setbox2=\vbox to 2 truein{\epsfxsize=4 in\epsfbox[0 0 612 792]{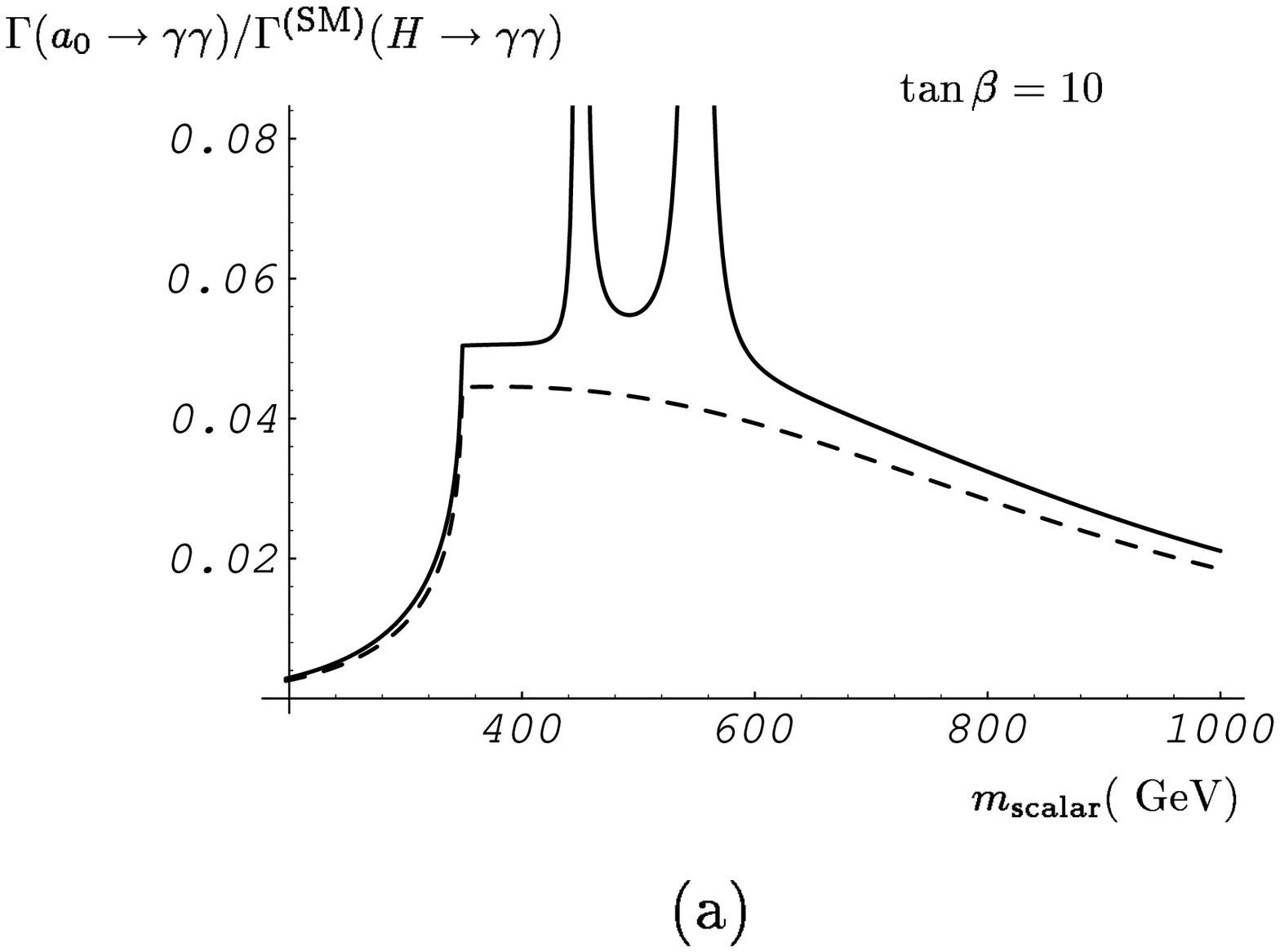}}
\setbox3=\vbox to 2 truein{\epsfxsize=4 in\epsfbox[0 0 612 792]{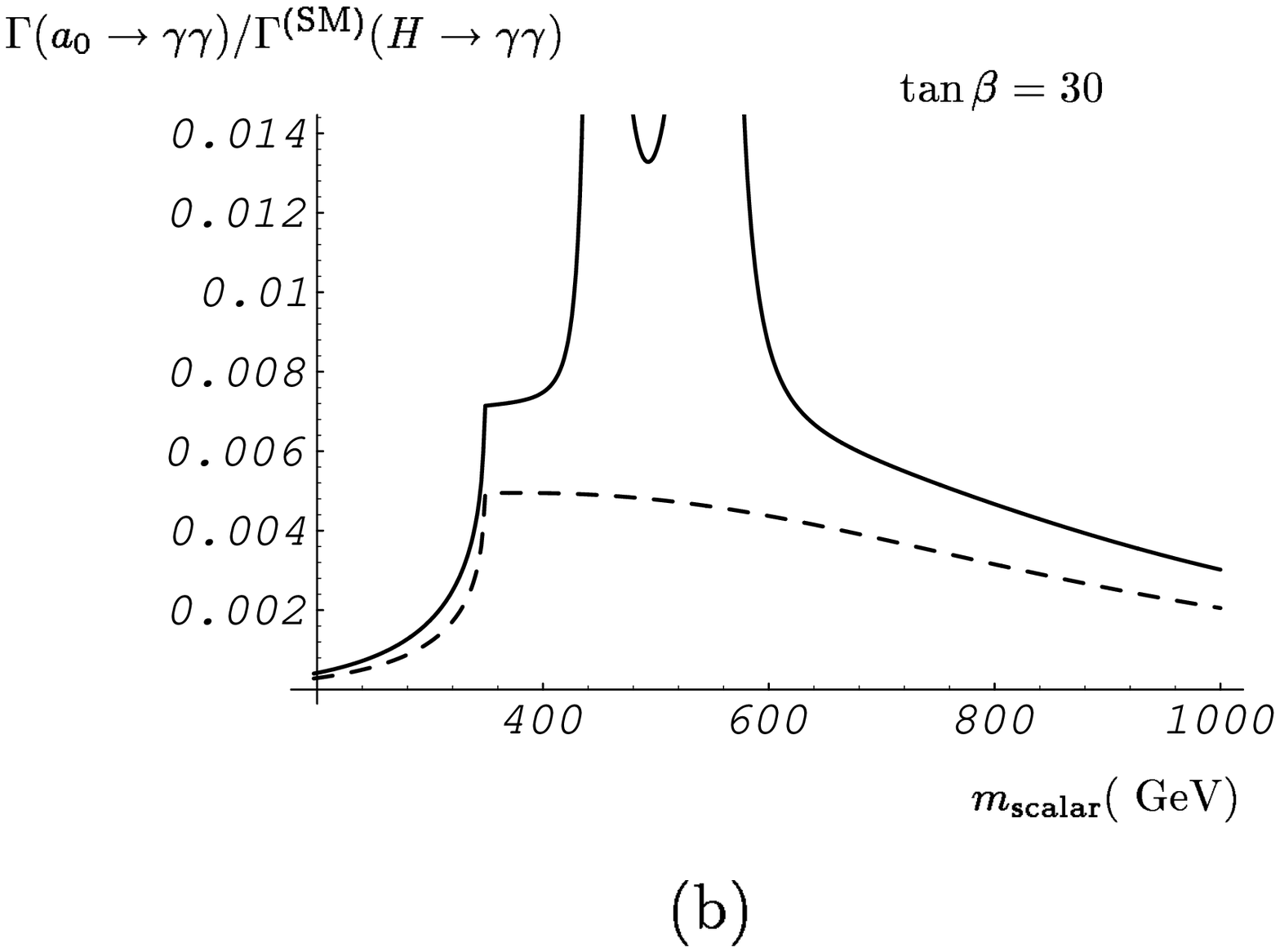}}
\setbox4=\vbox to 2 truein{\epsfxsize=4 in\epsfbox[0 0 612 792]{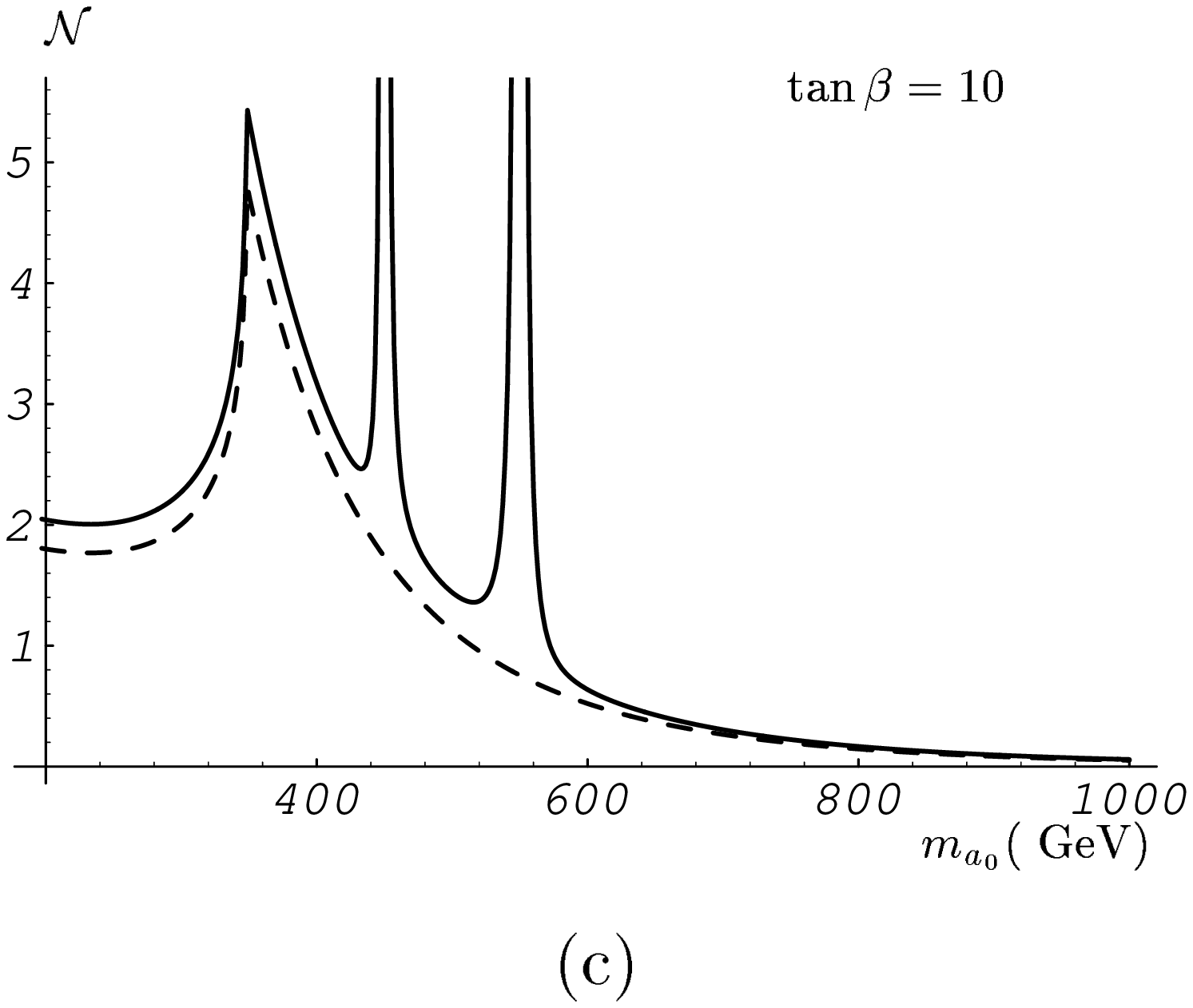}}
\setbox5=\vbox to 2 truein{\epsfxsize=4 in\epsfbox[0 0 612 792]{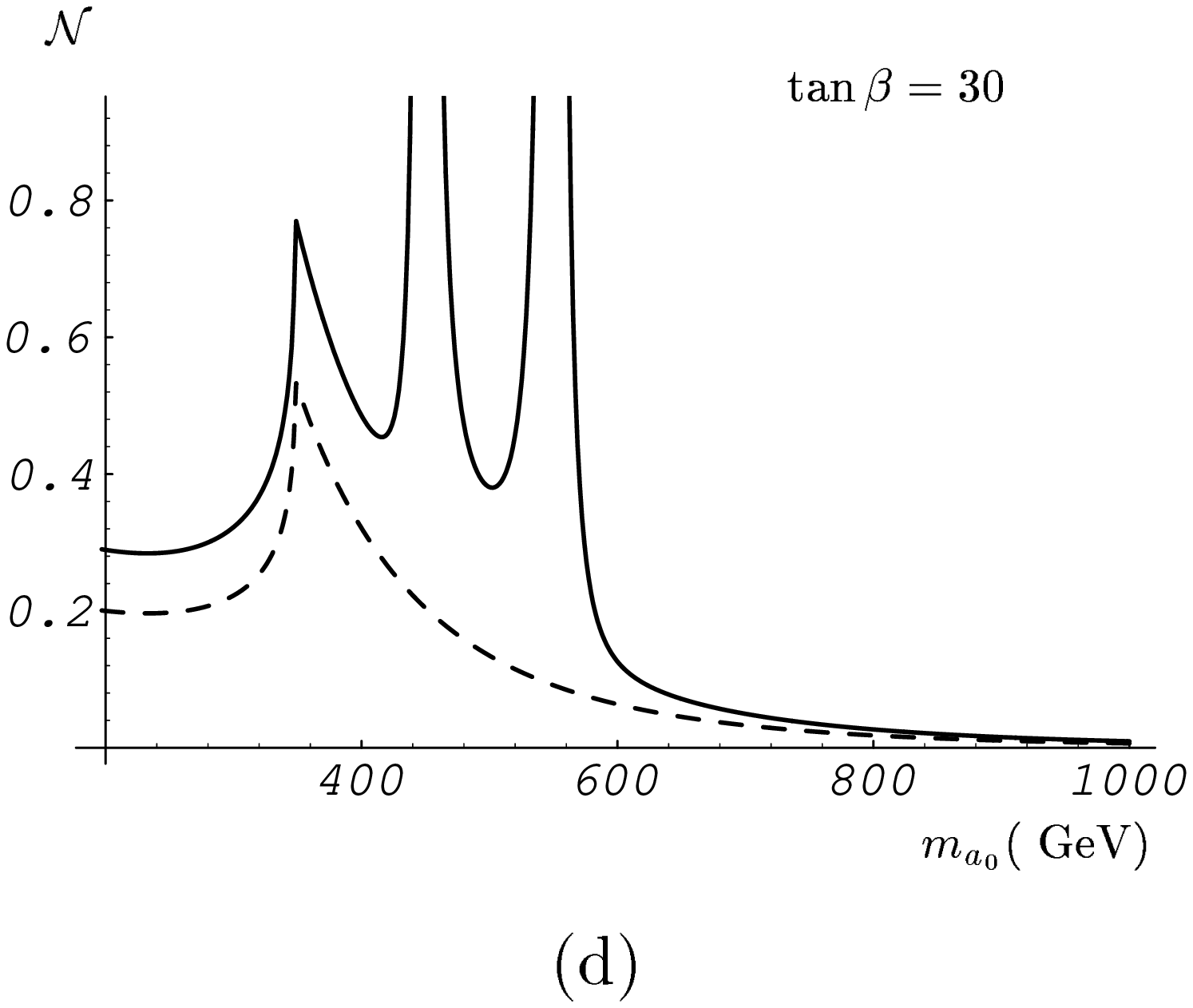}}

\vskip -25 pt

\line{ $\mskip-80mu$ \box2 $\mskip-40mu$ \box3 }

\bigskip\bigskip

\line{ $\mskip-60mu$ \box4 $\mskip-40mu$ \box5 }

\vskip 60pt

\thecaption{Figure 2.}{ {\bf(a),(b)}: Ratio of the total $ \ao \rightarrow
\gamma \gamma $ decay width
to the decay width of a \sm\ Higgs particle of the same mass into two
photons. {\bf(c),(d)} Total number of yearly events at the LHC (C.M.
energy$=7$TeV,
luminosity $100$/fb).
The solid curves
correspond to the theory with effective operators for $ \Lambda = 1 \tev $,
the dashed curves give
the pure two-doublet model results. We chose $ m_\ho = 450 \gev $, $
m_\hho = 550 \gev $, $ \mh = 500 \gev $ and $ \Lambda = 1 \tev $;
$ m \lowti{top}=174 \gev $, $ \alpha=45^o$.}

\bigskip

As can be seen from this figure there is a $ \sim50\%$ increase from
the light-physics values for this decay width; despite this one has to face
the complications produced by the smallness of the branching ratio,
being, in the most favorable case $ \sim  10^{-7 } $ and about 3\%\ of
the two-photon width of a  \sm\ Higgs boson of the same mass,
as shown in Fig. \widthratio a and \widthratio b. The number of events,
showed in \widthratio c,d is marginal for one year of LHC running, but,
mainly due to the resonance enhancement, deviations from the two doublet
model would be observable for moderate values of $\mao $.

This implies that when (and if) the  $ \ao $ is discovered
at the LHC, its branching ratio into two photons could prove a sensitive
reaction in which to study physics beyond this excitation but the
experimental sensitivity required is enormous. We have
verified that all heavy physics effects are completely obscured when $
\tan \beta $ is smaller than $ \sim 5 $; the results are also
insensitive to the mass of the charged Higgs.

We also investigated the behaviour of the two-photon branching ratio as
a function of $ \Lambda $; the results are presented in Fig. \lamdep.
Based on this graph we conclude that this reaction will be able to probe
physics up to $ \sim 1 \tev $.

\setbox2=\vbox to 2 truein{\epsfxsize=4.5 in\epsfbox[0 0 612 792]{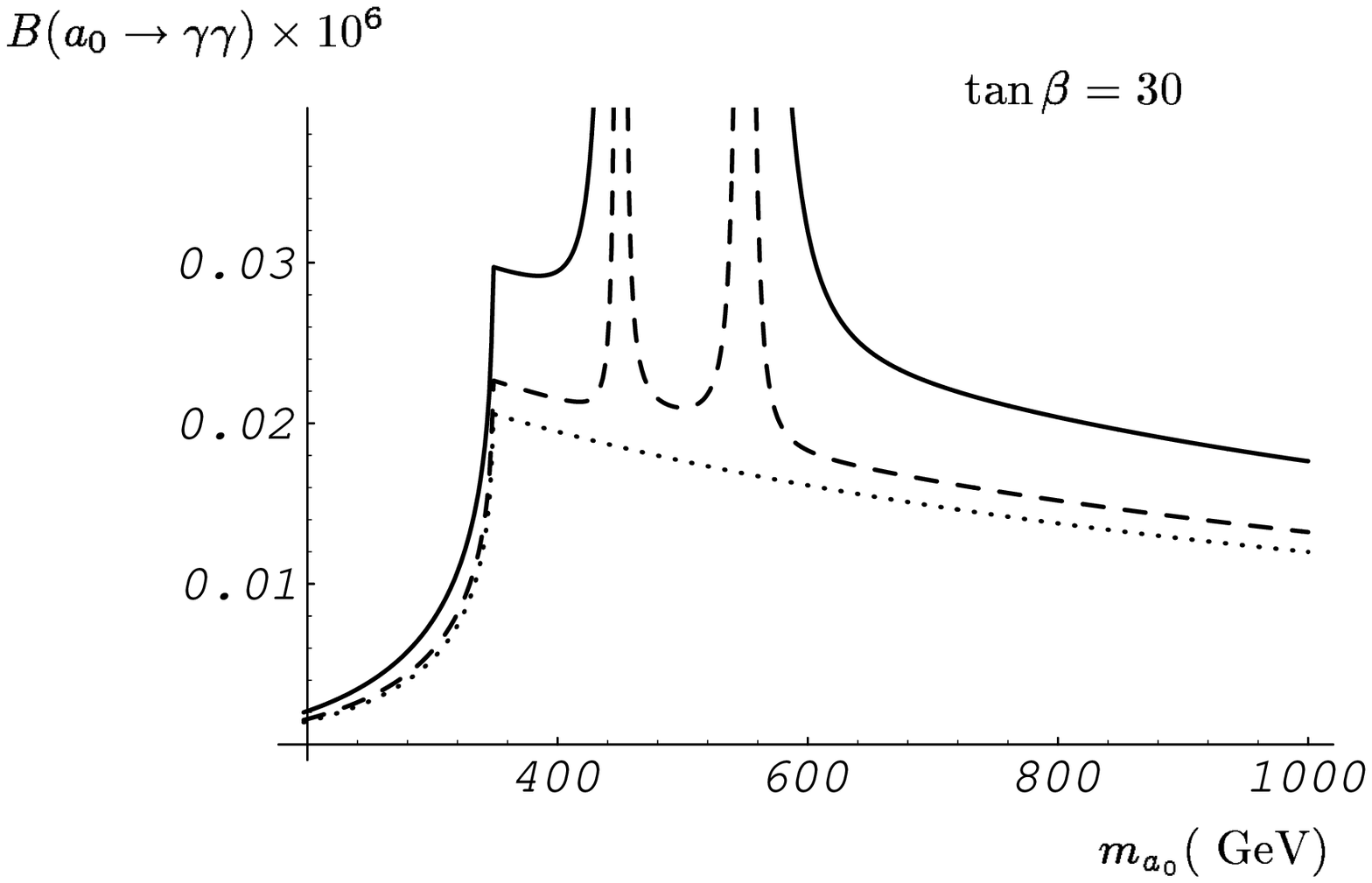}}

\vskip -10 pt

\centerline{ \box2  }

\bigskip\bigskip

\thecaption{Figure 3.}{ Branching ratio for the decay $ \aogg $ for $ \Lambda =
1
\tev $ (solid curve), $ \Lambda = 2 \tev $ (dashed curve) and $ \Lambda =
\infty $ (dotted curve). We chose $ m_\ho = 450 \gev $, $ m_\hho = 550
\gev $, $ \mh = 500 \gev $ and $ \tan\beta = 30 $;
$ m \lowti{top}=174 \gev $, $ \alpha=45^o$. }

\bigskip

Now we consider $ \ao $ production in photon-photon collisions.
 From general considerations the complete amplitude is of the form
$ \acal_{ \gamma \gamma \rightarrow \ao }
 = \mcal_1 \ppi + \mcal_2 \ppii $. We will assume that
both initial photons have the same polarization matrix. In this case we
obtain $$ | \acal_{ \gamma \gamma \rightarrow \ao } |^2 = { s^2 \over 8 }
 \left[ \left| \mcal_1 \right |^2 \left( 1 + \xibf _- \cdot \xibf_+\right)
+ \left| \mcal_2 \right |^2 \left( 1 - \xibf_+^2 \right) \right] \eqn\eq $$
where $ \xibf_\pm= ( \xi_1 , \pm \xi_2 , \xi_3 ) $ and $ \xi_i $
are the Stokes parameters for the photons (assumed the same for both).

The ratio $ ( 1 - \xibf_+^2) / ( 1 - \xibf_- \cdot \xibf_+ ) $ is independent
of the
polarization of the intial photons and depends only on the polarization of
the initial electrons (for perfectly polarized electrons $ \xibf^2_+ = 1 $).
As an illustration we will assume that the initial electrons are 70\%\
longitudinally polarized, the electron and photon energies are
taken equal to $ 500 \gev $ and $ 1.17 \ev $. In this case [\gamgam]
$ ( 1 - \xibf_+^2) / ( 1 - \xibf_- \cdot \xibf_+ ) \simeq 0.86 $
(and becomes $ \simeq0.7$ at 90\%\ polarization)~\foot{
For a more significant
reduction exquisite degrees of polarization are required: 99\% for
$ ( 1 - \xibf_+^2) / ( 1 - \xibf_- \cdot \xibf_+ ) \simeq 0.25 $
}.

\setbox2=\vbox to 2 truein{\epsfxsize=4 in\epsfbox[0 0 612 792]{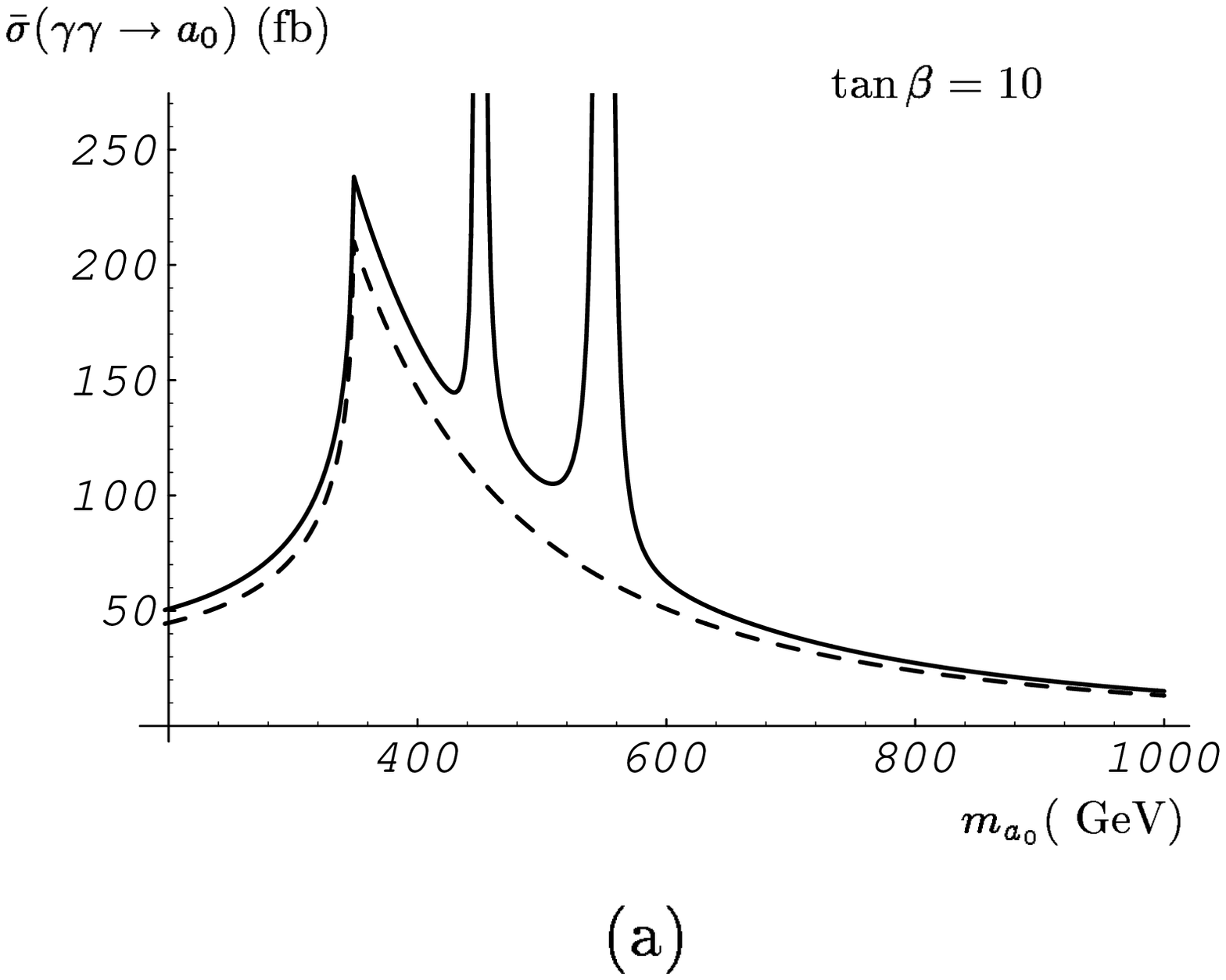}}
\setbox3=\vbox to 2 truein{\epsfxsize=4 in\epsfbox[0 0 612 792]{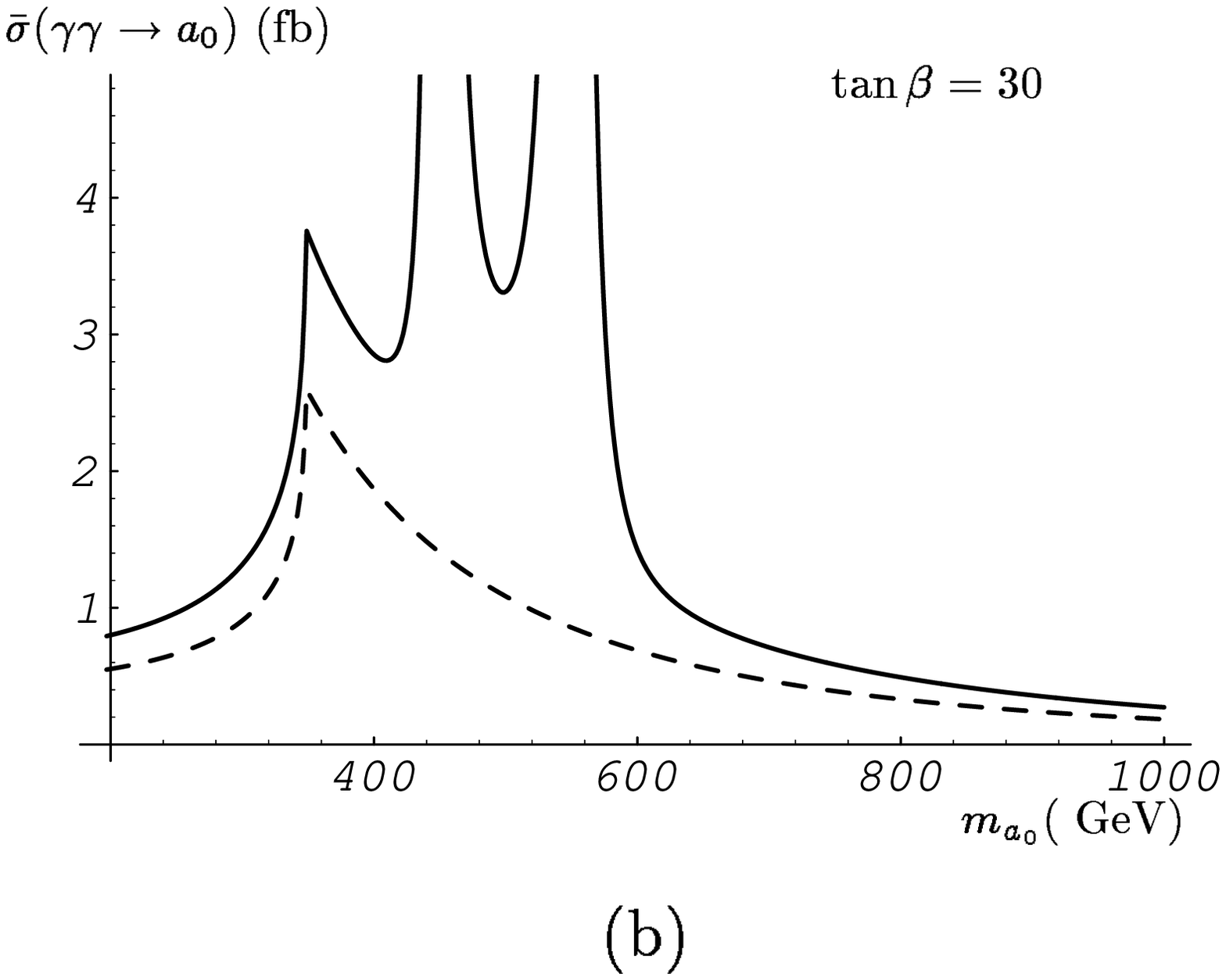}}

\vskip -10pt

\line{ $\mskip-80mu$ \box2 $\mskip-40mu$ \box3  }

\bigskip\bigskip

\thecaption{Figure 4.}{Total $ \ao $ production cross section averaged over an
energy bin equal to the total width of the $ \ao $. The solid curves
correspond to the theory with effective operators for $ \Lambda = 1 \tev $,
the dashed curves give
the pure two-doublet model results.
 We chose $ m_\ho = 450 \gev $, $
m_\hho = 550 \gev $, $ \mh = 500 \gev $;
$ m \lowti{top}=174 \gev $, $ \alpha=45^o$. The
electron polarization is 70\% .}

\bigskip

 These results imply
that, even though it is theoretically possible to eliminate
the light-physics contribution to this cross section for any value
of $ \beta $, experimental constraints on the initial electron's
degree of polarization prevent the realization of this possibility.
The expression for the cross section is simply (in the center of mass
system) $$ d \sigma = { \pi \over 2 s \; \mao }
\left| \acal_{ \gamma \gamma \rightarrow \ao } \right|^2 \delta \left(
\sqrt{ s } - \mao \right ) . \eqn\eq $$

In figure \csect\
we plot the value of this
expression averaged over the width of the $ \ao $, namely
 $$ \bar \sigma = { \pi \over 2  \mao^3 }
{ \left| \acal_{ \gamma \gamma \rightarrow \ao } \right|^2 \over
\Gamma_\ao }. \eqn\eq $$

To illustrate the significance of this result
we consider the ratio of $ \bar \sigma $ to the same quantity when $
\Lambda \rightarrow \infty $. As can be seen from Fig. \ratio, the
deviations are important for $ \Lambda = 1 \tev $.

\setbox2=\vbox to 2 truein{\epsfxsize=4.5 in\epsfbox[0 0 612 792]{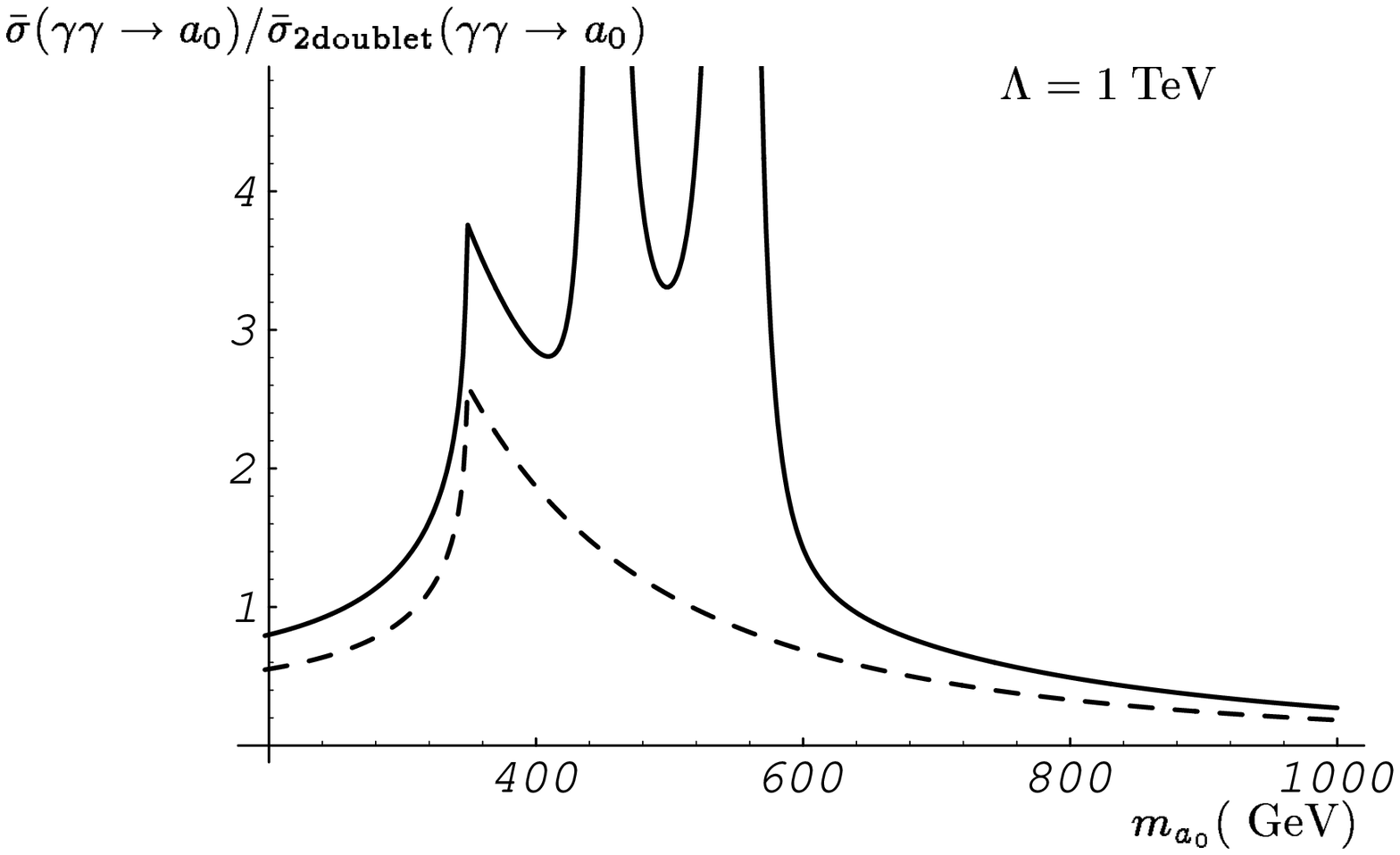}}

\centerline{ \box2  }

\vskip 30 pt

\thecaption{Figure 5.}{ Ratio of the total to light cross sections. The solid
curve
correspond to $ \tan \beta = 30$, the dashed curve to $ \tan\beta = 10 $.
We chose $ m_\ho = 450 \gev $, $
m_\hho = 550 \gev $, $ \mh = 500 \gev $ and $ \Lambda = 1 \tev $;
$ m \lowti{top}=174 \gev $, $ \alpha=45^o$. The
electron polarization is 70\% .}

\bigskip

The number of events to be generated at a $ \gamma \gamma
$ collider with $ 10 /$fb$/$yr luminosity is presented in Fig. 6. As in
the previous cases, deviations from the usual two-doublet model are
noticeable for large values of $ \tan\beta $ only.

\setbox2=\vbox to 2 truein{\epsfxsize=4 in\epsfbox[0 0 612 792]{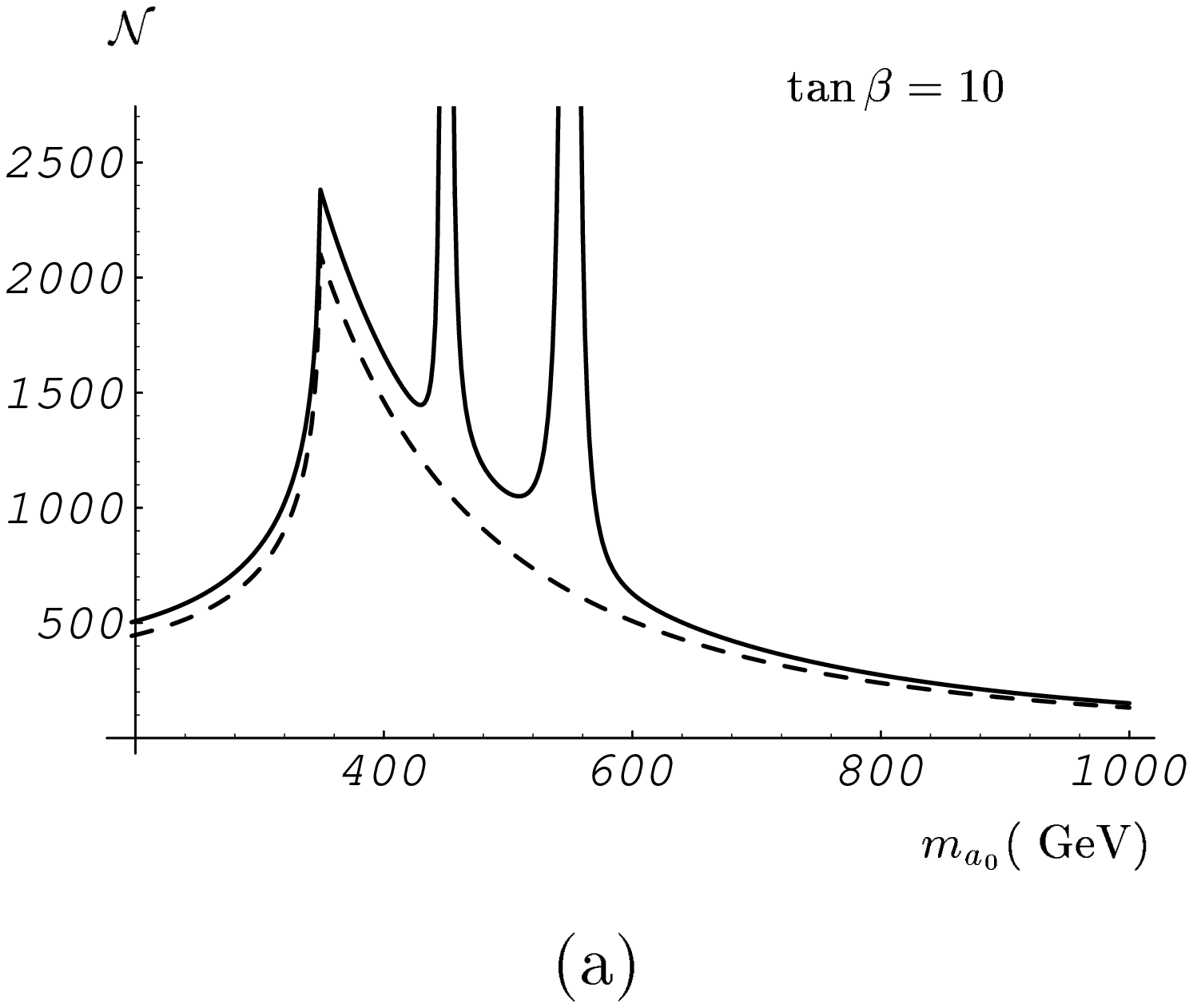}}
\setbox3=\vbox to 2 truein{\epsfxsize=4 in\epsfbox[0 0 612 792]{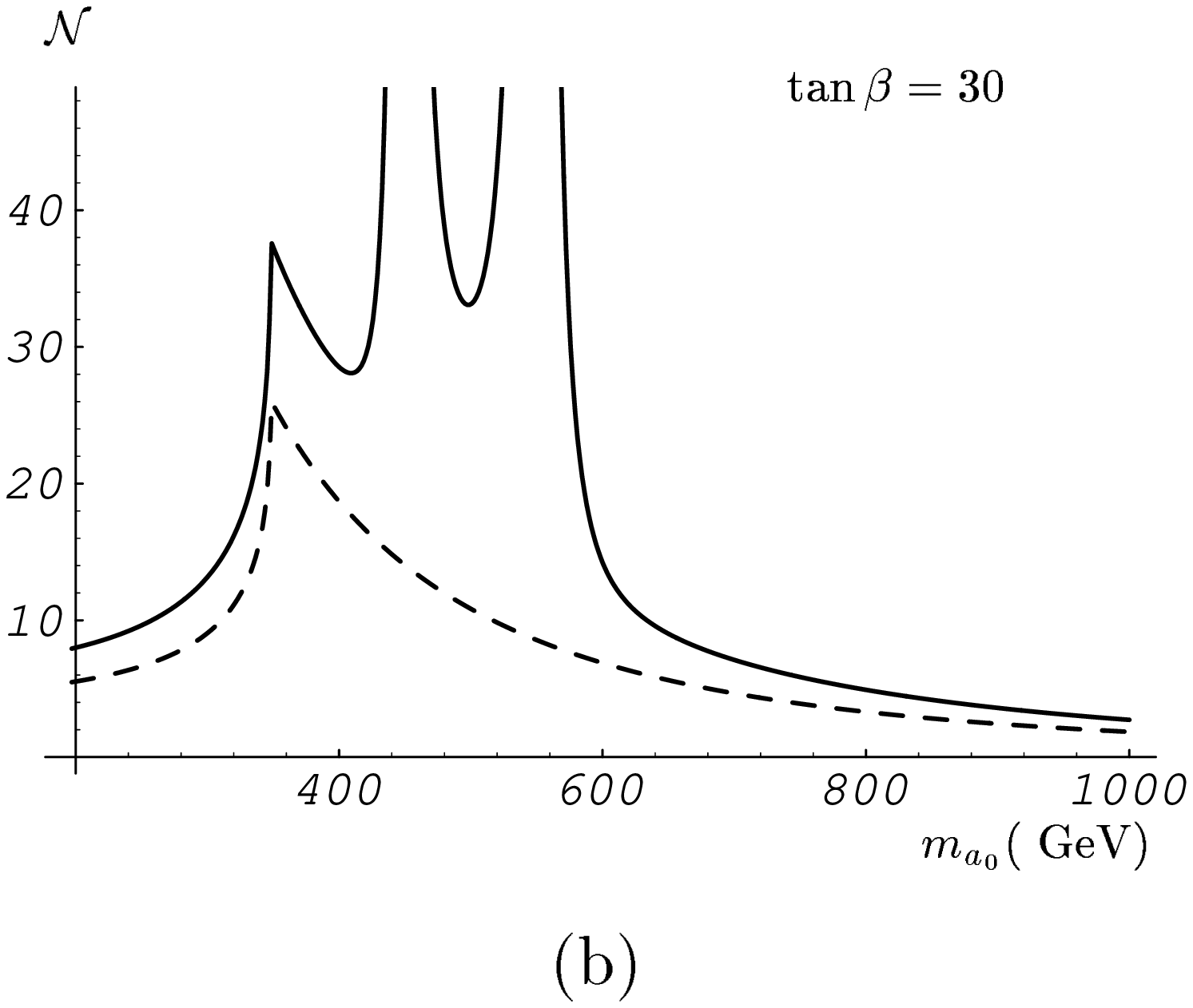}}


\line{ $\mskip-80mu$ \box2 $\mskip-40mu$ \box3  }
\vskip 30 pt
\thecaption{Figure 6.}{Number of $ \ao $ events produced in
$ \gamma \gamma $ colissions for two values of $ \tan\beta $.
We chose $ m_\ho = 450 \gev $, $
m_\hho = 550 \gev $, $ \mao = \mh = 500 \gev $;
$ m \lowti{top}=174 \gev $, $ \alpha=45^o$. The
electron polarization is 70\%  and the luminosity equals 10/fb.}

\setbox2=\vbox to 2 truein{\epsfxsize=4 in\epsfbox[0 0 612 792]{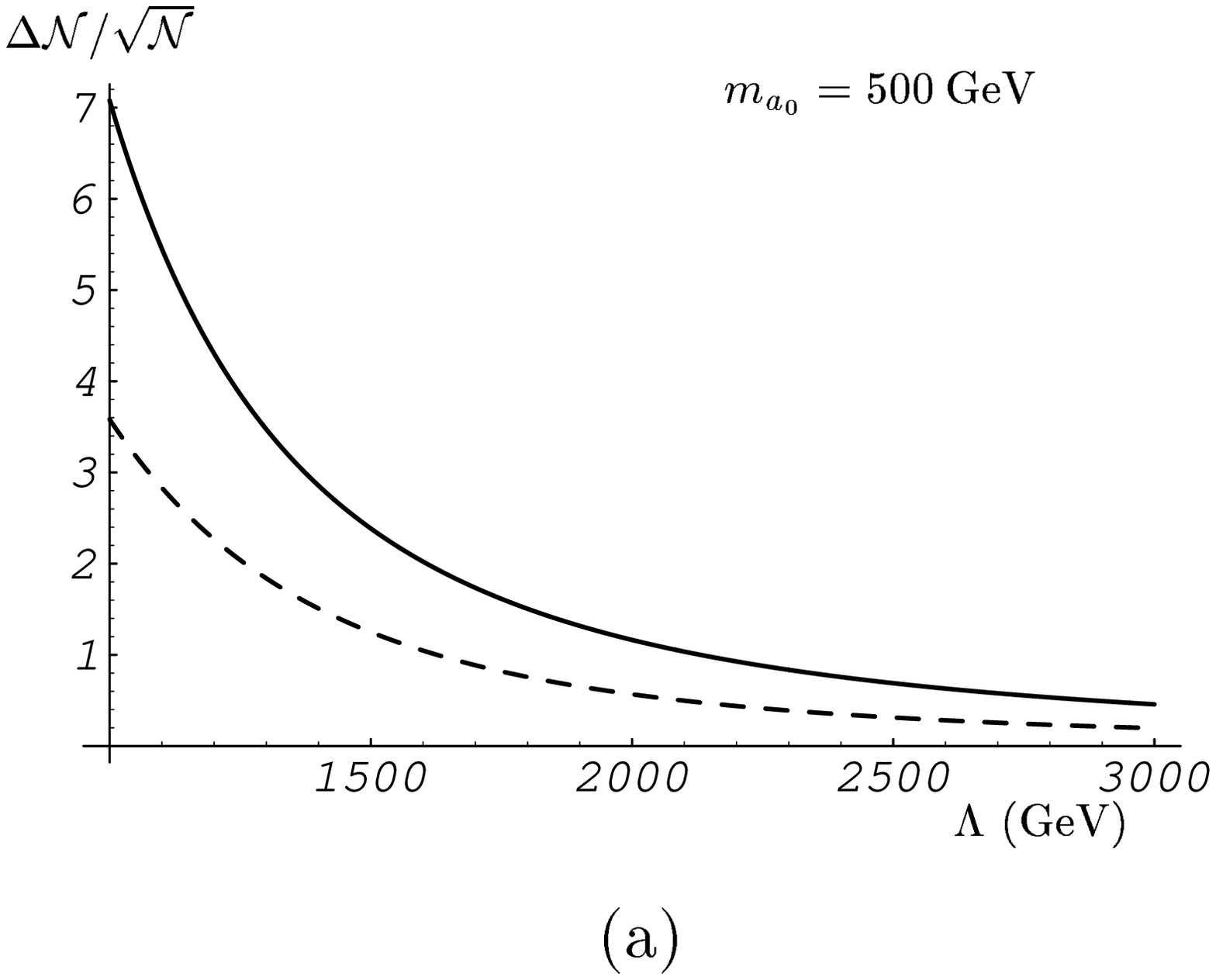}}
\setbox3=\vbox to 2 truein{\epsfxsize=4 in\epsfbox[0 0 612 792]{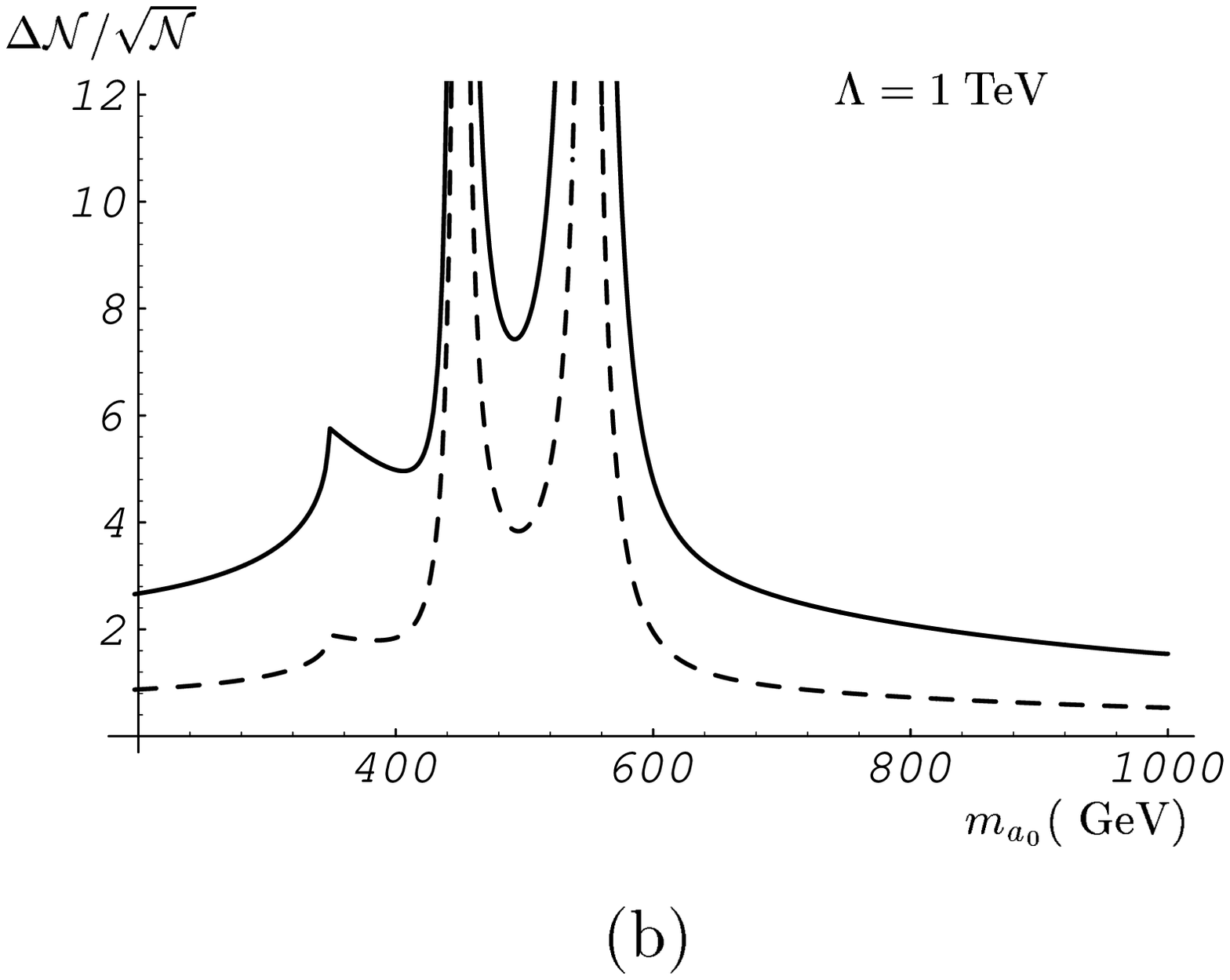}}


\line{ $\mskip-80mu$ \box2 $\mskip-40mu$ \box3  }

\vskip 40 pt

\thecaption{Figure 7.}{Statistical significance of the deviations
from the light cross section,
solid curve $ \tan \beta = 30 $, dashed curve
$ \tan\beta = 10 $. We chose $ m_\ho = 450 \gev $, $
m_\hho = 550 \gev $, $ \mao = \mh = 500 \gev $;
$ m \lowti{top}=174 \gev $, $ \alpha=45^o$. The
electron polarization is 70\%  and the luminosity equals 10/fb.}


Finally, in order
to estimate the statistical significance of the deviations from the
light-physics result, we evaluate $ \ncal $, the total number of events and
$ \Delta \ncal $, the total number of events minus the number of events
present when $ \Lambda \rightarrow \infty $. Then $ \Delta \ncal / \sqrt{
\ncal } $ is a measure of the statistical significance of the deviations
from the light-physics results and is plotted in Fig. \statsign\ for a 10/fb
luminosity. As can be seen from this plot there are significant
deviations from the light-physics results up to a few \tev.

\chap{ Conclusions}

We have employed the method of effective lagrangians in evaluating the
possible deviations from the two-doublet extension of the standard model
for reactions containing the CP odd excitation $ \ao $ and two photons.
The calculations illustrate the fact that effective lagrangians can be used in
loop calculations.

The magnitude of the resets can be estimated reliably within a set of
general scenarios. For example, we must examine whether CP violation
operators are generated at the same scale as the CP conserving ones, or whether
the underlying physics is weakly or strongly coupled. One of the advantages
of the effective lagrangian approach is the ease with which these various
possibilities are identified and studied.

The cases which we studied in detail were the two-photon processes
related to the CP-odd scalar $ \ao $. We found important deviations
stemming from the anomalous couplings when the scale of new physics is
moderate large (a few \tev) provided $ \tan \beta \gesim 20 $. There
are several problems with the type of processes considered stemming
mainly from the small branching ratios and cross sections. For the LHC,
the event number is so reduced (see Fig. 2) that the observation of
the effects generated by the physics beyond the two-doublet model is
very unlikely (except for a 10-year study). In constrast, for a $ \gamma
\gamma $ collider with a  luminosity of $\sim$10/fb, a careful study
will uncover some effects generated by the physics beyond
the two doublet model, or at least place significant bounds on the scale
of this type of new physics (see Figs. 6,7.).

It was mentioned in the introduction (as
evident from the form of the potential) that no CP violation effects
were included in the light theory. In contrast, several effective
operators (such as $ \ocal \up{1,3}_{ i j k l }$) violate CP. This situation
is not inconsistent. In fact, CP violations in the potential can be
pictured as $ \ho - \ao $ and $ \hho - \ao $ mixings of the same type as
those induced by $ \ocal \up{1 , 3 } $. Therefore the presence of CP
violating terms in the potential will not affect the conclusions as long as
their
order of magnitude is the same as that of the mixing terms generated by the
effective operators. More precisely, we have assumed that CP violation
is generated at a scale $ \gesim \Lambda $.

The numerical results were obtained not by the most optimistic choice of
parameters, but by mimicking the possible presence of cancellations
among various graphs. As noted in the appendix there is still the
possibility of having further suppression factors, in which case the new
physics effects will be too small to be observed. On the other hand
there could be some enhancements, in which case our results would
constitute a lower bound on the new physics effects.

The results presented in this paper are to be compared
to the ones derived in the minimal \sm\ where
the (CP-even) $H \gamma \gamma $ interaction
is generated through one-loop effects of charged fermions and
$W$ gauge bosons. The
contributions from the $W$ boson and the top quark are dominant, although
the latter is only marginally important and tends to cancel the first one
partially.
%
%
When the contribution of dimension six operators is considered, the
virtual effects may enhance the \sm\ width of this rare decay mode up to
one order of magnitude.~\refmark{\derujula,\contreras} Recently it was
pointed out that tree-level generated bosonic operators of dimension
eight may also enhance the \sm\ result.~\refmark\contreras
%

The conclusions of this paper are, by necessity, speculative. The $ \ao $
has not been found as of yet and so the study of its decays lies in the
future. Still, based on the above calculations it is clear that if this
excitation is observed, the detailed study of it's interactions with two
photons may open a window into new physics.

\ack

The authors would like to thank the UC Mexus program for providing the
funds for this research. J.W. is grateful to E. Ma and M.B. Einhorn for
interesting conversations and for the support provided by the SSC
fellowship FCFY9211. M.A.P. and J.J.T. acknowledge financial support
from CONACyT (M\'exico).

\appendix

\baselineskip 19 pt

In this appendix we present the full list of operators contributing to the
$ \aogg $ three point function and the corresponding
amplitudes.

The list of (tree-level-generated)
operators which contribute at one loop is the following ($ i , k , l , m
, n = 1 , 2 $)
$$\eqalign{
& \ocal_{ e \varphi ; i j k } = \! \left( \phi_i^\dagger \phi_j \right) \!\!
      \left( \bar \ell e \phi_k \right) ; \cr &
\ocal_{ d \varphi ; i j k } = \! \left( \phi_i^\dagger \phi_j \right) \!\!
         \left( \bar q d \phi_k \right) ; \cr &
\ocal_{ u \varphi ; i j k } = \! \left( \phi_i^\dagger \phi_j \right) \! \!
  \left( \bar q u \tilde \phi_k \right) ; \cr
&\ocal_{ \phi_k f } =  i \left( \phi_k^\dagger D_\mu \phi_k^{} \right)
                  \left( \bar f \gamma^\mu f \right) ; \quad f = e , u , d ;
\cr
&\ocal_{ \phi_k F } \up 1 = i \left( \phi_k^\dagger D_\mu \phi_k \right)
                             \left( \bar F \gamma^\mu F \right) ; \quad F =
\ell , q ; \cr
&\ocal_{ \phi_k F } \up 3 = i \left( \phi_k^\dagger \tau^I
                        D_\mu \phi_k \right)
                    \left( \bar \ell \gamma^\mu \tau^I \ell \right) ; \quad F =
\ell , q ; \cr
&\ocal_{ \varphi; i j k l m n } = \left( \phi^\dagger_i \phi_j \right)
                                  \left( \phi^\dagger_k \phi_l \right)
                                  \left( \phi^\dagger_m \phi_n \right) ;
\cr &
\ocal_{ \partial \varphi ; i j k l  } = \half
           \left[ \partial_\mu \left( \phi^\dagger_i \phi_j \right) \right ]
           \left[ \partial_\mu \left( \phi^\dagger_k \phi_k \right) \right ] ;
 \cr  &
\ocal_{ i j k l } \up 1 =
               \left[ \partial_\mu \left( \phi^\dagger_i \phi_j
\right) \right] \left[ \phi_k^\dagger { \buildrel \leftrightarrow
\over D_\mu } \phi_l \right] ;  \cr &
\ocal_{ i j k l } \up 3 =
               \left[  \phi^\dagger_i D_\mu \phi_j
\right] \left[ \left( D^\mu \phi_k \right)^\dagger
\phi_l \right] ; \cr} \eqn\olist$$  where only operators
consistent with the discrete symmetry are retained, for example $
\ocal_{ e \varphi; i j k } $ is considered only with $ i + j + k $
odd, $ \ocal_{ \varphi ; i j k l m n } $ when $ i + j + k + l + m + n $
even, etc. As mentioned above, we have adopted the notation of Ref.
\bw. It is worth pointing out
that the operator $ \ocal \up 1 _{ i j k l } $ is slightly different
from the one used in Eq. (3.14) of Ref.\bw, the difference between this
expression and ours is easily seen to be, using the equations of
motion, a linear combination of $ \ocal_{ \varphi ; i j k l m } $ and
$ \ocal_{ \partial \varphi ; i j k l } $; therefore both expressions
are equivalent [\effeom]. The one we chose enormously simplifies the
calculation
of the corresponding amplitude (reducing contributions from 19 to 3
graphs).

The operators $ \ocal_{ \varphi; i j k l m n } , \
\ocal_{ \partial \varphi ; i j k l  } $ and
$ \ocal_{ i j k l } \up { 1 , 3 }  $, generate quadratic terms which
mix $ \ao $ with the scalars, $
h^0 $ and $ H^0 $ (this is a consequence of the fact that
these operators violate CP). Therefore these operators produce
two kind of graphs in the unitary gauge: 1PI diagrams with $H^+$
in the loop, and 1PR diagrams where the $ \ao $ converts into a
$ h^0$ or $ H^0$ via the above mixing terms, and then the
$ h^0$ or $ H^0 $ decay into two photons via $H^+$, $W$ or fermion
loops. The amplitude corresponding to the above $ \ocal $ via 1PR graphs
will be denotes by a ``$(1PR)$'' superscript (no 1PI superscript
will be used in the other amplitudes in order to simplify the notation)

We denote by $ \acal_i $ the value of the on-shell three point
function generated by $ \ocal_i $ (without the factor $ y_i / \Lambda^2
$), straightforward evaluation of the
relevant Feynman rules and graphs yields, for the 1PI contributions containing
bosons in the loop, generated by $ \ocal_{ \varphi i j k l m } ,
\ocal_{ \partial \varphi i j k l } $ and $ \ocal \up a _{ i j k l } , \
( a = 1 , 3 ) $ yield ($f$ stands for the right-handed
fermions $ f = e , u , d$)
$$ \eqalign{
\left. \matrix{ \acal_{ \varphi k k 1 2 1 2 }  \cr
                \acal_{ \varphi k k 2 1 2 1 }  \cr} \right\}
=& \pm { ( - )^k \mw^3 \sw^3 s_{ 4 \beta } \over
4 \pi^2 \mao^2 e } \left[ 2 { \mh^2 \over \mao^2 } I_{ - 1 } ( \mh )
+ 1 \right] \ppi \cr
\left. \matrix{\acal_{ \partial \varphi 1 2 1 2} {} \cr
               \acal_{ \partial \varphi 2 1 2 1} {} \cr} \right\}
=& \mp {\mw \sw e s_{ 2 \beta } \over 8 \pi^2 } \left[
2 { \mh^2 \over \mao^2 } I_{ - 1 } ( \mh ) + 1 \right] \ppi \cr
\acal_{ i j k l } \up a {} {} =&
 { e\; \mw \sw s_{ 2 \beta } h \up a _{ i j k l } \over 4 \pi^2 }
\left[ 2 { \mh^2 \over \mao^2 } I_{ - 1 } ( \mh ) + 1 \right] \ppi;
\qquad a = 1 , 3 \cr }
\eqn\ascalars $$
where $ \mh $ denotes the mass and $$
I_n ( \mu ) = \int_0^1 dx \, x^n \ln \left[ 1 - x ( 1 - x )
\mao^2 /\mu^2 \right]  ; \qquad n \ge -1 . \eqn\eq $$
The coefficients in \ascalars\ are
$$ \matrix{
h \up 1 _{ 11kk } =  ( - )^k \sb^2 ; \hfill &
h \up 1 _{ 22kk } = ( - )^k \cb^2 ; \hfill &
h \up 1 _{ 1212 , 1221 , 2112 , 2121 } = - \half c_{ 2 \beta } . \hfill \cr
& & \cr
h \up 3 _{ 2211 , 2121 } = +1 ; \hfill &
h \up 3 _{ 1122 , 1212 } = -1 ; \hfill &
h \up 3 _{ 1111 , 2222 , 1221 , 2112 } = 0 . \hfill \cr }
\eqn\eq $$

The 1PI contributions with fermions in the loop are
$$ \eqalign{
\acal_{f \varphi i j k } {} =& { \sb b^f_{ i j k } \mw^2 \sw^2
Q_f^2 \mf \over 2 \sii \, \pi^2 \mao^2 }
\Biggl\{ \left[ \left( { 4 \mf^2 \over \mao^2 } - 1 \right) I_{ - 1 } ( \mf )
+ 2 \right] \ppi \cr & \mskip340mu - I_{ -1 } ( \mf ) \ppii \Biggr\} \cr
\acal_{ \phi_k f } {} =&{ e \sw \mw s_{2 \beta }  Q_f^2
( - ) ^k \over 8 \pi^2 }
\left[ 2 { \mf^2 \over \mao^2 } I_{ - 1 } ( \mf ) +1 \right]  \ppii \cr
\acal \up{ 1 , 3 }_{ \phi_k , \ell  } {}
=& { e \sw \mw s_{2 \beta } Q_f^2  ( - )^{ k + 1 } \over 8 \pi^2 }
\left[ 2 { \me^2 \over \mao^2 } I_{ - 1 } ( \me ) +1 \right]  \ppii \cr
\acal \up{ 1 , 3 }_{ \phi_k , q  } {}
=& { e \sw \mw s_{2 \beta } Q_f^2  ( - )^{ k + 1 } \over 8 \pi^2 }
\sum_{ f = d , u } d_f\up{ 1 , 3 }
\left[ 2 { \mf^2 \over \mao^2 } I_{ - 1 } ( \mf ) +1 \right]  \ppii \cr
} \eqn\eq $$
where
$$ \eqalign{
& b^{ e , d } _{ 111 , 221 , 212 , 122 } =
-\cb^2 , -\sb^2 , - \sb^2 , 1 + \cb^2 ; \quad
b^u_{ 112 , 121 , 211 } = \cb^2 , \cb^2 , - 1 - \sb^2 \cr
& d_f\up a  = \cases{ - 1 & for $ a = 3 , \ f = u $ \cr \
                      + 1 & otherwise \cr } \cr
%
} \eqn\eq $$

Finally the 1PR contributions are generated, as mentioned above,
by the contact terms in some operators which induce $ \ao - h^0 $
and $ \ao - H^0 $ mixings, the $ h^0$ and $H^0$ then decay into
two photons via fermion, $W$ or $H^+$ loops (in the unitary gauge).
The results can be extracted from [\okun, \hhg], the amplitude
is ($ \phi = h^0 , \ H^0 $) $$ \eqalign{
\acal_\phi \up{1PR} ( \ocal_i ) = &
{ \qwe_i ( \phi )   \over \mao^2 - m_\phi^2 }
{ e \sw \mw^3 \over 2 \pi^2  }
\Biggl\{  u_{ \phi H }  \left[ 1 +
         { 2 \mh^2 \over \mao^2 } I_{ - 1 } ( \mh ) \right] \cr
& + 3  u_{ \phi W }  \left\{
\left({ \mao^2 \over 2 \mw^2 } - 1 \right) \left[ 1 +
{ 2 \mw^2 \over \mao^2 } I_{ - 1 } ( \mw ) \right] + \half \right\} \cr
& + \sum_f Q_f^2 N_f
{ \mf^2 u_{ \phi f } \over 4 \mw^2 } \left[
\left( 1 - { 4 \mf^2 \over \mao^2 } \right) I_{ - 1 } ( \mf )
- 2 \right] \Biggr\} \ppi \cr } \eqn\eq $$
where $N_f $ denotes the number of colors for fermion $f$.
The non-vanishing constants $ \eta_i(\phi) $ are
$$ \qwe (h^0) \up1 _{ 11kk} =  (-)^{ k + 1 } \cb^2 \sb \sa ; \qquad
\qwe (h^0) \up1 _{ 22kk}  = (-)^k \sb^2 \cb \ca ; $$
$$ \qwe (h^0) \up1_{1212,2121}  = { 3 \camb + \captb \over 4 } ; \qquad
\qwe (h^0) \up1_{1221,2112}  = - \sb \cb \sapb $$
$$ \qwe (H^0) \up1 _{ 11kk} = (-)^k \cb^2 \sb \ca ; \qquad
\qwe (H^0) \up1 _{ 22kk} = (-)^k \sb^2 \cb \sa ; $$
$$\qwe (H^0) \up1_{1212,2121} = { 3 \samb + \saptb \over 4 } ; \qquad
\qwe (H^0) \up1_{1221,2112} = \sb \cb \capb $$
$$
\qwe (h^0) \up3_{1122,2121} = - \qwe  (h^0) \up 3 _{ 2211 , 1212} =
                       \quarter s_{2 \beta } \samb $$
$$
\qwe (H^0) \up3_{1122,2121} = - \qwe (H^0) \up 3 _{ 2211 , 1212} = -
                       \quarter s_{2 \beta } \camb $$
$$ \qwe ( h^0 )_{ \partial \varphi ; 1212 }  = -
\qwe( h^0 )_{ \partial \varphi ; 2121 } = { 3 \over 2 } \camb - \half \ctb
\camb $$
$$ \qwe ( H^0 )_{ \partial \varphi ; 1212 }  = -
\qwe ( H^0 )_{ \partial \varphi ; 2121 } = { 3 \over 2 } \samb - \half \ctb
\samb $$
$$\eqalign{ &\qwe ( h^0) _{ \varphi ; k k 1 2 1 2 } = -
\qwe ( h^0) _{ \varphi ; k k 2 1 2 1 } = { \sw^2 \mw^2
\left( 1 - ( - )^k \ctb \right) \over 4 \pi \alpha
\mao^2 }  \Biggl[ { 5 \over 2 } \capb
\cr & \qquad \qquad \qquad \qquad \qquad \qquad \qquad \qquad \qquad
+ \camb \left( ( - 1 )^k - \half \ctb \right) \Biggr] \cr
& \qwe ( H^0)_{ \varphi ; k k 1 2 1 2 }  = -
\qwe ( H^0)_{ \varphi ; k k 2 1 2 1 }  = { \sw^2 \mw^2
\left( 1 - ( - )^k \ctb \right) \over 4 \pi \alpha
\mao^2 }  \Biggl[ { 5 \over 2 } \sapb
\cr & \qquad \qquad \qquad \qquad \qquad \qquad \qquad \qquad \qquad
-  \samb \left( ( - 1 )^k - \half \ctb \right) \Biggr] \cr
} \eqn\eq $$
and
$$ \matrix{
& u_{ \phi H } & u_{ \phi W } & u_{ \phi u } & u_{ \phi d } \cr
( \phi=h^0 ) & - \samb  + { \ctb \sapb \over 2 \cw^2 } & - \samb & \ca / \sb
& - \sa / \cb \cr
( \phi=H^0 ) & \camb - { \ctb \capb \over 2 \cw^2 } & \camb & \sa / \sb
& \ca / \cb \cr } \eqn\eq $$

The total contribution is
then
$$ \acal = \acal \lowti{standard} + \inv{\Lambda^2 }
\left( \sum_i \al_i \acal_i
+ \sum_i \al_i \acal_i \up{1PR} \right) \eqn\eq $$
where the constants $ y_i$ contain all couplings
from  the underlying theory which generates the operators $ \ocal_i $.
 From this expression the decay width of the $ \ao $ or its production cross
section can be evaluated directly. Note that the terms containing $ \ppi $
and $ \ppii $ will not interfere in polarization averaged widths or
cross-sections.

It must be kept in mind that all the coefficients will contain, in general,
some coupling constants of the underlying theory which are small
($ \lesim 1 $) by our assumption of it being a weakly coupled theory.
For illustrative purposes we will choose $ | \al_i | = 1 $
and we will neglect all anomalous
contributions which are proportional to a fermion mass
other than the top. In order to simulate possible cancellations between
the various contributions of the same type (produces by constants $
\al_i $ of opposite signs) we will use an averaging procedure such as
$$ \sum_{ i j k = 112 , 121 , 211 }
y_{ t \varphi ; i j k } \acal_{ t \varphi ; i j k } \rightarrow
\inv 3
\sum_{ i j k = 112 , 121 , 211 }  \acal_{ t \varphi ; i j k } \eqn\eq $$
This simple procedure implies several cancellations, for example,
the contributions from $ \acal_{ \phi_k f } $ and $ \acal
\up{ 1 , 3 } _{ \phi_k f} $ cancel.

We emphasize that this example is presented
for illustrative purposes only, nonetheless we expect the results to provide
good semi-quantitative estimates. Should there be fewer cancellations
the anomalous signal would be enhanced; in this case the results
presented constitute a lower bound. On the other hand it is possible for
the constants $ y_i $ could be suppressed by unknown effects in which
case our results would over-estimate the anomalous effects.

\refout\expel
\figout
\bye